\documentclass{sig-alternate}
\usepackage{mathptmx} 

\usepackage[normalem]{ulem}
\usepackage[hyphens]{url}
\usepackage[sort,nocompress]{cite}
\usepackage[final]{microtype}
\usepackage[keeplastbox]{flushend}
\usepackage{hyperref}
\usepackage{xspace}
\usepackage{cite}
\usepackage{amsmath,amssymb,amsfonts}
\usepackage[format=plain,font=small,bf]{caption}
\usepackage{subcaption}
\usepackage{algorithmic}
\usepackage{graphicx}
\usepackage{graphics}
\usepackage{fancyhdr}
\usepackage{textcomp}
\usepackage[table, dvipsnamesxcdraw]{xcolor}
\usepackage{cite}
\usepackage{url}
\usepackage[normalem]{ulem}
\usepackage{verbatim}
\usepackage{listings}
\usepackage{enumitem}
\usepackage{hhline}
\usepackage{threeparttable}
\usepackage{colortbl}
\def\BibTeX{{\rm B\kern-.05em{\sc i\kern-.025em b}\kern-.08em
    T\kern-.1667em\lower.7ex\hbox{E}\kern-.125emX}}
    
\usepackage{tabularx}
\usepackage{multirow,array}
\usepackage{booktabs}
\newcommand{\microsubmissionnumber}{422}

\usepackage{authblk}
\usepackage{multicol}

\fancypagestyle{firstpage}{
  \fancyhf{}
  
  \fancyhead[C]{\vspace{15pt}\normalsize{MICRO 2020 Submission
      \textbf{\#\microsubmissionnumber} -- Confidential Draft -- Do NOT Distribute!!}} 
  \fancyfoot[C]{\thepage}
}

\newcommand{\ignore}[1]{}
\newcommand{\system}{\textit{Sesame}\xspace}

\newcommand{\ckmarkl}{$\checkmark$}
\newcommand{\crossl}{$\times$}
\newcommand{\ckmarkd}{$\checkmark$}
\newcommand{\crossd}{$\times$}
\newcommand{\RCG}[1]{\rowcolor[HTML]{656565}}
\newcommand{\RCC}[1]{\rowcolor[gray]{0.9}}
\definecolor{darkgrey}{HTML}{434343}
\newcommand{\mrow}[2][c]{%
      \begin{tabular}[#1]{@{}c@{}}#2\end{tabular}}
\newcommand{\WT}[1]{\color{white}}
\newcommand{\RC}{\cellcolor[gray]{0.9}}

\newcommand{\prakash}[1]{{\color{purple}[#1]}}

\newcommand{\revision}[1]{{\color{black}#1}}
\usepackage[utf8]{inputenc}

\definecolor{codegreen}{rgb}{0,0.6,0}
\definecolor{codegray}{rgb}{0.5,0.5,0.5}
\definecolor{codepurple}{rgb}{0.58,0,0.82}
\definecolor{backcolour}{rgb}{0.95,0.95,0.92}

\lstdefinestyle{mystyle}{
    commentstyle=\color{codegreen},
    keywordstyle=\color{magenta},
    emph=[1]{LOAD,GEMM,STORE},emphstyle=[1]{\color{blue}\bfseries},
    emph=[2]{LOAD_E,STORE_E,GEMM_C,ZEROIZE},emphstyle=[2]{\color{magenta}\bfseries},
    numberstyle=\tiny\color{codegray},
    stringstyle=\color{codepurple},
    basicstyle=\ttfamily\tiny,
    breakatwhitespace=false,         
    breaklines=true,                 
    captionpos=b,                    
    keepspaces=true,                 
    numbersep=3pt,
    frame=none,
    keepspaces=true,
    showspaces=false,                
    showstringspaces=false,
    showtabs=false,                  
    tabsize=2
}
\usepackage{caption}
\usepackage{enumitem}

\newcounter{nalg}
\renewcommand{\thenalg}{\arabic{nalg}} 
\DeclareCaptionLabelFormat{algocaption}{Pseudocode \thenalg} 

\lstnewenvironment{algorithm}[1][] 
{   
    \refstepcounter{nalg} 
    \captionsetup{labelformat=algocaption,labelsep=colon} 
    \lstset{
        mathescape=true,
        frame=tB,
        numbers=left, 
        numberstyle=\tiny,
        basicstyle=\ttfamily\footnotesize, 
        keywordstyle=\color{purple},
        keywords={input, output, return, datatype, function, in, if, else, foreach, while, begin, end, for},
        numbers=left,
        xleftmargin=.04\textwidth,
        #1 
    }
}
{}

\title{SESAME: \b{S}oftware defined \b{E}nclaves to \b{S}ecure Inference \b{A}ccelerators with \b{M}ulti-tenant \b{E}xecution}
\author[1]{Sarbartha Banerjee}
\author[2]{Prakash Ramrakhyani}
\author[1]{Shijia Wei}
\author[1]{Mohit Tiwari}
\affil[1]{University of Texas at Austin}
\affil[2]{ARM Research}
\affil[ ]{\textit {\{sarbartha,shijiawei\}@utexas.edu\quad prakash.ramrakhyani@arm.com\quad tiwari@austin.utexas.edu}}

\begin{document}

\pagestyle{plain}
\maketitle

\begin{abstract}
\quad Hardware-enclaves that target complex CPU designs 
compromise both security and performance.
Programs have little control over micro-architecture, which leads to
side-channel leaks, and then have to be transformed to have worst-case 
control- and data-flow behaviors and thus incur considerable slowdown.
We propose to address these security and performance problems by
bringing enclaves into the realm of accelerator-rich architectures. \\
\indent The key idea is to construct \textit{software-defined enclaves} (SDEs)
where the protections and slowdown are tied to an 
application-defined threat model and tuned by a compiler for the
accelerator's specific domain. This vertically integrated approach
requires new hardware data-structures to partition, clear, and shape
the utilization of hardware resources; and a compiler that instantiates and schedules these data-structures to create multi-tenant enclaves on accelerators. \\
\indent We demonstrate our ideas with  a comprehensive prototype -- \system{} -- that includes modifications to compiler, ISA, and microarchitecture to a decoupled access execute (DAE) accelerator 
framework for deep learning models. 
Our security evaluation shows that classifiers that could
distinguish different layers in VGG, ResNet, and AlexNet, fail to do so when run using \system{}. Our synthesizable hardware prototype 
(on a Xilinx Pynq board) demonstrates how the
compiler and micro-architecture enables threat-model-specific trade-offs in code size increase ranging from 3-7 $\%$ and run-time performance overhead for specific defenses ranging from 3.96$\%$ to 34.87$\%$ (across confidential inputs and models and single vs. multi-tenant systems).

\end{abstract}

\section{Introduction}

Hardware-enclaves are key to confidential computing~\cite{confidential-computing-consortium} -- where users can push their private data into a box that is invisible to privileged software, co-resident processes, and even attackers with physical access to the pins of the chip~\cite{aegis, xom, bonsai-merkle-tree,ascend, Phantom,SGX,sanctuary,Sanctum,keystone}. Confidential computing can provide a trustworthy foundation where users can safely work with healthcare, financial, or business data while organizations can offload compliance enforcement~\cite{gdpr, ccpa} to services that provide hardware root of trust.

Enclaves today however offer a Hobson's choice. A user can pick a design like Intel SGX that hardwires a very specific threat model for general-purpose CPUs -- and pay with performance-overheads~\cite{sgx-is-slow, Vault} and the risk of side-channel breaches~\cite{Foreshadow,RIDL,Fallout,attackSurvey} -- or be left with the default unprotected execution inside virtual machines. 
Crucially, confidential computing is limited to general-purpose cores while accelerator-rich architectures~\cite{TPU,TVM} worsen the gap between non-secure and enclaved execution for important classes of programs like deep learning and graph computing\cite{Graphicianado,gui2019survey}.

\ignore{
\prakash {Current hardware paradigms for secure computing like Intel SGX\cite{SGX} are riddled with some challenging problems. Firstly they tend to be seriously limited on the size of applications they can support before letting performance fall over the cliff \cite{Vault}. Secondly, they are mostly focused on securing applications running on a general purpose CPUs. Thirdly, there are a myriad of micro-architectural side-channel attacks\cite{RIDL,Foreshadow, Fallout,attackSurvey} requiring the need of a software-hardware co-design approach while designing TEEs. These CPUs have predominantly been made with the one overarching design goal of improving performance, or (more lately) improving performance within a given power budget and even later to address the security vulnerabilities in design}. The solution space to addressing the security vulnerabilities is tricky due to the vastness of attack surface and one can easily venture into solutions\cite{invisispec,specshield,STT} that require heavy if not prohibitive performance costs and/or end up fixing only part of the problem. Domain-specific secure accelerators not only reduce the attack surface but also provide software-hardware solution opportunities to better address TEE designs.
}

\ignore{
\begin{itemize}
    \item TEE paradigm shift to domain-specific accelerators
    \item Full-stack accelerator enclaves
    \item configurable threat model<Performance and security tradeoffs>
    \item multi-tenant implementation
\end{itemize}
}

In this paper, we propose to bring confidential computing to domain-specific secure accelerators, retaining most of the performance benefits of using an accelerator (compared to CPUs) and also reducing the attack surface by closing side-channels.
The key insight is to enable \textit{software-defined enclaves} -- where software can construct enclaves that are customized for a specific threat-model and program domain -- in order to optimize performance without leaking secrets. Threat models are only known at deployment-time, and hard-wiring one into hardware means that every confidential program pays the price of security against threats they may not care about -- for example, the cost of integrity checks or oblivious main memory accesses in a secure data-center facility, or the price of obfuscating code when it might be public. 
Similarly, general purpose program execution have to be obfuscated assuming worst-case control- and data-flows and uncontrolled micro-architectural side-effects~\cite{raccoon,escort} which incurs significant slowdowns. Software-defined enclaves can tune the slowdowns from security to scale gracefully with threat models while driving security and performance optimizations down from algorithms to bits and gates.


Specifically, we introduce \system{}, a software-defined enclave framework for multi-tenant machine learning inference accelerators that are tightly coupled to a CPU (e.g., Arm Ethos-N NPUs \cite{ArmEthos-N,Arm-ML}). We focus on decoupled access-execute (DAE) architectures as an example accelerator that is popular across deep learning, graph, and other data-driven domains across cloud and edge devices. \system{} takes deep learning models such as VGG, ResNet, and AlexNet as input and produces as output an auto-tuned program and a multi-tenant hardware design that enforces computational non-interference between security domains. Crucially, a user can express their threat model in terms of (a) the program model and/or the user inputs being secret, and (b) whether the attacker's visibility includes on-chip or off-chip signals. \system{} translates all concurrent users' threat models into a lattice of security labels and ensures computational non-interference among all labels -- (informally) ensuring that secret inputs/model information from one user does not leak via on-chip and/or off-chip signals to other users (including the cloud provider or hardware-owner).

\system{} includes defense mechanisms across the layers of the accelerator stack. 
At the synthesizable hardware (RTL) level, we introduce two new data-structures that the \system{} compiler can instantiate and tune for the current threat model and program: (1) \textit{private queue} that software can partition based on a schedule -- these queues prevent \textit{contention-channel leaks} and replace queues that are ubiquitous in hardware designs between pipelined stages and arbiters of shared hardware; (2) \textit{traffic shaper} that decouples \textit{observation channels} -- where attackers infer secrets from signals coming out of a software-defined enclave -- from secret variables. While partitioning and shaping are generic strategies that have been applied from networking~\cite{vuvuzela} to hardware~\cite{camouflage}, \system{}'s contribution is to enable compiler/synthesis tools to overlay non-secure RTL designs with private queues and shapers and re-use the data movement logic to create flexible enclaves. Further, for traffic shapers that can be software-configured, their RTL implementation can be far simpler than (e.g.) hardware-only shapers designed specifically to obfuscate main memory traffic~\cite{camouflage}.  

At the compiler level, \system{}'s auto-tuning phase generates a tiling schedule that maximizes performance while ensuring that the computation and memory access schedule has no secret information. The code-generator then annotates instructions appropriately to obfuscate observation channels like execution variability and based on the threat model the driver turns on defenses like private queues and traffic shaping to quash contention and external observation channels. 

Enabling a constant-time mode for execution units (like GEMM and ALU) is the last piece required to construct multi-tenant enclaves -- our prototype system's baseline GEMM and ALU units happen to be constant-time already and hence we do not use this feature.

\ignore{While these ideas have been studied in isolation before, e.g., Camouflage~\cite{camouflage} to shape memory traffic, \system{} composes these hardware data-structures and compiler passes to provide strong secrecy guarantees during accelerator execution even in the presence of a strong enclave-model adversary that can observe all digital signals on- and off-chip (we discuss details in the threat model in Section~\ref{sec:ThreatModel}).}

\ignore{Prototype information? Attack para? Or -- can leave it to bullets below}

To summarize, we make the following contributions: 
 \begin{itemize}[leftmargin=*]
    \item \system{} introduces software-defined enclaves for domain specific accelerators, enabling enclave software to be confidential against co-tenants and infrastructure providers.
    
    \item A detailed vulnerability analysis of on-chip accelerators, highlighting gaps that need defenses in order to construct a range of accelerator enclaves. 
    \item A cross-layer design including new hardware data-structures to shape traffic and to share queues (in addition to using standard base-bound techniques to partition on-chip state); exposed via ISA extensions to software; that a compiler uses to generate code for a user-specified threat model. 
    
    \item An end-to-end implementation of a deep-learning accelerator (using a baseline VTA~\cite{VTA} system) where \system{} compiles and synthesizes six workloads -- including ResNet, VGG, and AlexNet models -- to a Xilinx FPGA.
    
    \item A security evaluation that shows a classifier that could distinguish different layers in each model (the first step towards reversing models and weights) is foiled by \system{}; and performance evaluation in terms of code-size, slowdown, and area cost of \system{} across a range of system configurations and threat models. For example, slowdown varies from $3.96\%$ to $29.2\%$ under increasingly confidential and contention-heavy settings.
    
 \end{itemize}

We will open source our hardware and compiler contributions to 
spur further research into software-defined enclaves for DAE-applications beyond deep learning.
~\\

\begin {comment}
Interestingly, domain-specific programs often have simpler semantics -- e.g., there are no pointers in matrix-computation based deep learning models -- which simplifies compiler analysis.

\end{comment}

\section{Background}
\label{sec:background}

\subsection{Baseline Secure Platform}
\label{sec:baseline-security}
Our baseline consists of a user on a client system who engages one or more services deployed by an infrastructure provider on a cloud system. The infrastructure provider can expose accelerator resources in standard units (e.g., small/medium/large instances on Amazon's F1 cloud) that the user's \system{} compiler can generate code for. Alternatively, a high level model may be communicated securely to the cloud which is then compiled in a sandbox to ensure no information is leaked. The binary generated is securely transferred to the provisioning service which copies it into accelerator memory or uses it as a bitstream to configure an FPGA.   

To program an accelerator, the provisioning service has to create a trusted channel to establish a root of trust on the accelerator. This entails the following assumptions about such a platform:
\begin {itemize}[leftmargin=*]
    \itemsep0em
\item The platform provides an attestation service to assure the client that
\begin{itemize}
    \itemsep0em
\item All hardware on the platform is provided by a trusted manufacturer. This ensures the platform is free from any hardware trojans with accelerator bitstream hash checked during secure boot.
\item The platform is capable of deploying a trusted execution environment (TEE). This includes the ability to isolate an application from the OS, hypervisor and other privileged/non-privileged software on both the CPU and main memory. Further, this TEE can be extended to provide similar isolation guarantees for the system memory used by the accelerator.
\item The ML compiler framework and driver are sourced from a trusted software provider and can be deployed to run in a TEE.
\end{itemize}
Attestation protocols remain the same whether a `secure processor' implements a RISC CPU (like Aegis ~\cite{aegis} or Sanctum~\cite{Sanctum}) or an application-specific accelerator logic. Attestation enables platform authentication (whose identity is vouched for by a public-key certificate authority) on which the user can execute a workload 

\item The host platform and the client have the ability to set up a secure channel to communicate sensitive information
\item Confidential data stored in main memory can be protected using encryption \cite{AMD-SME,AWS-Graviton2,Intel-TME}.
\item Any key management systems that handle cryptographic keys to either set up communication channels or provision memory encryption for the accelerator are in the trusted code base. 
\end{itemize}

\subsection {Baseline Accelerator Architecture}
VTA\cite{VTA} is a DAE\cite{DAE} machine built on the TVM\cite{TVM} software stack to provide a domain-specific end-to-end solution to neural network applications. It is built on a Xilinx Zynq-7000 series FPGA. Users write their model and explicitly schedule the computation into a high-level CISC representation. This is then converted into low-level accelerator instructions by a code-generator running in an FPGA co-processor. The accelerator then uses these instructions and data to perform inference in the accelerator programmed in the FPGA. \par

\section{Vulnerability Analysis}
\label{sec:VulAnal}

\begin{table*}[t!]
\centering
\small
\begin{tabular}{|p{0.4cm}|p{4.5cm}|p{4.2cm}|>{\centering}m{0.7cm}|>{\centering}m{0.7cm}|p{5.0cm}|}
  \rowcolor[HTML]{656565}
  \multicolumn{1}{c|}{} & & &\multicolumn{2}{c|}{\color{white}\textbf{Threat Model}} & \\ 
  \cline{1-3}\cline{6-6}
  
  \rowcolor[HTML]{656565}
  \multicolumn{1}{c|}{\multirow{-2}{*}{\color{white}\textbf{No.}}} &
  \multicolumn{1}{c|}{\multirow{-2}{*}{\color{white}\textbf{Vulnerability}}} &
  \multicolumn{1}{c|}{\multirow{-2}{*}{\color{white}\textbf{Exploit}}} &
  \multicolumn{1}{c|}{\color{white}\textbf{PM}} &
  \multicolumn{1}{c|}{\color{white}\textbf{PI}} &
  \multicolumn{1}{c|}{\multirow{-2}{*}{\color{white}\textbf{Defences}}}\\
  \hline

    \rowcolor[HTML]{FFFFFF}
    1 & Adversary has access to communication channel between edge device and cloud  &
    Tampering of model/input in flight when being communicated to the accelerator &
    \ckmarkl & \ckmarkl & 
    Remote attestation of cloud system and establishing a secure channel using TLS \\ \hline
    
    \rowcolor[HTML]{FFFFFF}
    2 & Unauthorized access by privileged process on host CPU &
    Tampering of input/model in memory &
    \ckmarkl & \ckmarkl &
    Application isolation using a TEE framework like Aegis, Sanctum etc. \\ \hline

   \rowcolor[HTML]{FFFFFF}
    3 & Shared access to system memory carve-out by all tenants on accelerator and their host processes on CPU  &
    Adversary deploys an accelerator task to access victim's model parameters / input&
    \ckmarkl & \ckmarkl &
    TEE provides memory isolation between different tenants.  \\ \hline

    \rowcolor[HTML]{FFFFFF}
    4 & Compiler runs on the host system and has access to the model  &
     Model privacy compromised &
    \ckmarkl & \ckmarkl &
     Compiler is attested, made part of the trusted compute base and runs in a TEE.\\ \hline

    \rowcolor[HTML]{FFFFFF}
    5 & Model execution termination channel observation   &
     Model parameters may leaked due to runtime variability &
    \ckmarkl & \ckmarkl & 
     Scheduler allocates time slice increments at a granularity identified by model provider. \\ \hline
     
    \rowcolor[HTML]{B8D8ED}
    6 & Adversary has access to reading system memory &
    Secret data confidentiality violated  &
    \ckmarkl & \ckmarkl &
    Data encrypted by AES128 or QARMA128\\ \hline 

    \rowcolor[HTML]{B8D8ED}
    7 & Adversary has ability to observe memory bandwidth characteristics &
    Model topology and layer size leak &
    \ckmarkd & \crossd &
    Data independent distribution of memory traffic shape\\ \hline

    \rowcolor[HTML]{B8D8ED}
    8 & Shared access to dependency queues in accelerator &
    Loading corrupt data \& performing computation before data ready &
    \ckmarkd & \crossd &
    Partitioned Queues\\ \hline
    
    \rowcolor[HTML]{B8D8ED}
    9 & Shared access to instruction queues in accelerator &
    Instruction queue occupation serves as a covert channel &
    \ckmarkd & \crossd &
    Partitioned Queues with base $\&$ bound check\\ \hline
    \rowcolor[HTML]{B8D8ED}
    10 & Shared access to scratchpad in accelerator &
    Reading raw secret data on-chip by distrustful tenants &
    \ckmarkd & \ckmarkd &
    Base $\&$ bound check for each process and ZEROIZE after secret data becomes dead\\ \hline
    \rowcolor[HTML]{B8D8ED}
    11 & Shared access to execution units in accelerator &
    Sniffing execution output from the execution unit pipeline &
    \ckmarkd & \ckmarkd &
    Spatial partitioning of GEMM units with private operand buffers\\ \hline  
    
    \rowcolor[HTML]{F0F0F0}
    12 & Variable time GEMM unit execution &
    Model parameters / input data values leak &
    \ckmarkd & \ckmarkd & 
    Disabling data driven optimizations through GEMM.C instruction \\ \hline
    
    \rowcolor[HTML]{F0F0F0}
    13 & Adversary has ability to modify system memory &
    Secret model weights $\&$ filter maps can be tampered   &
    \ckmarkl & \ckmarkl  &
    Data MAC authentication\cite{MorphCount,mee,bonsai-merkle-tree} \\ \hline 

    \rowcolor[HTML]{F0F0F0}
    14 & Adversary has ability to observe addresses access patterns on memory bus &
    Memory access pattern attacks\cite{ReverseCNN} &
    \ckmarkl & \crossl &
     Invisimem\cite{Invismem},  PathORAM\cite{PathORM}\\ \hline

 \end{tabular}
\caption{\textbf{Vulnerabilities, exploits and defence mechanisms. Vulnerability 1-5 are addressed in the baseline platform, 6-11 are \system{} contribution and 12-14 can be composed with \system{}.  PM = Private Model, PI = Private Input.}}
  \label{tab:VulAnal}
  \vspace{-1em}
\end{table*}

 We systematize the vulnerabilities that a machine-learning-as-a-service(MLaaS) service may be subject to in Table~\ref{tab:VulAnal}. Vulnerabilities (1-5) are addressed in our baseline secure platform model. Vulnerabilities (6-11) correspond to the digital side channels arising within the accelerator that \system{} addresses.
In this section we take a closer look at these vulnerabilities and their applicability under different threat model variants relevant to \system{}. 

\begin{figure}[t]
    \centering
    \begin{subfigure}{\linewidth}
        \centerline{\includegraphics[width=\linewidth]{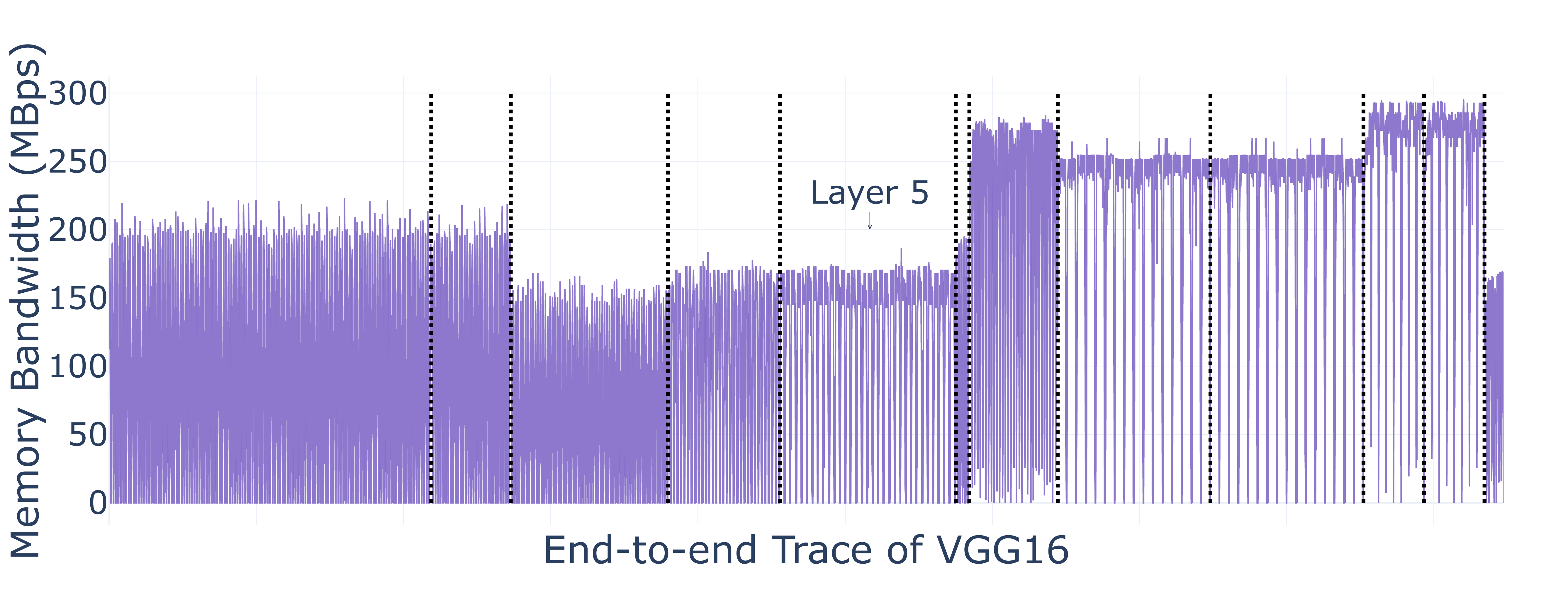}}
        \vspace{-0.5em}
        \subcaption{\label{fig:vgg16_end2end}\textbf{Memory bandwidth of layers in vgg16.}}
    \end{subfigure}
    \begin{subfigure}{\linewidth}
        \centerline{\includegraphics[width=\linewidth]{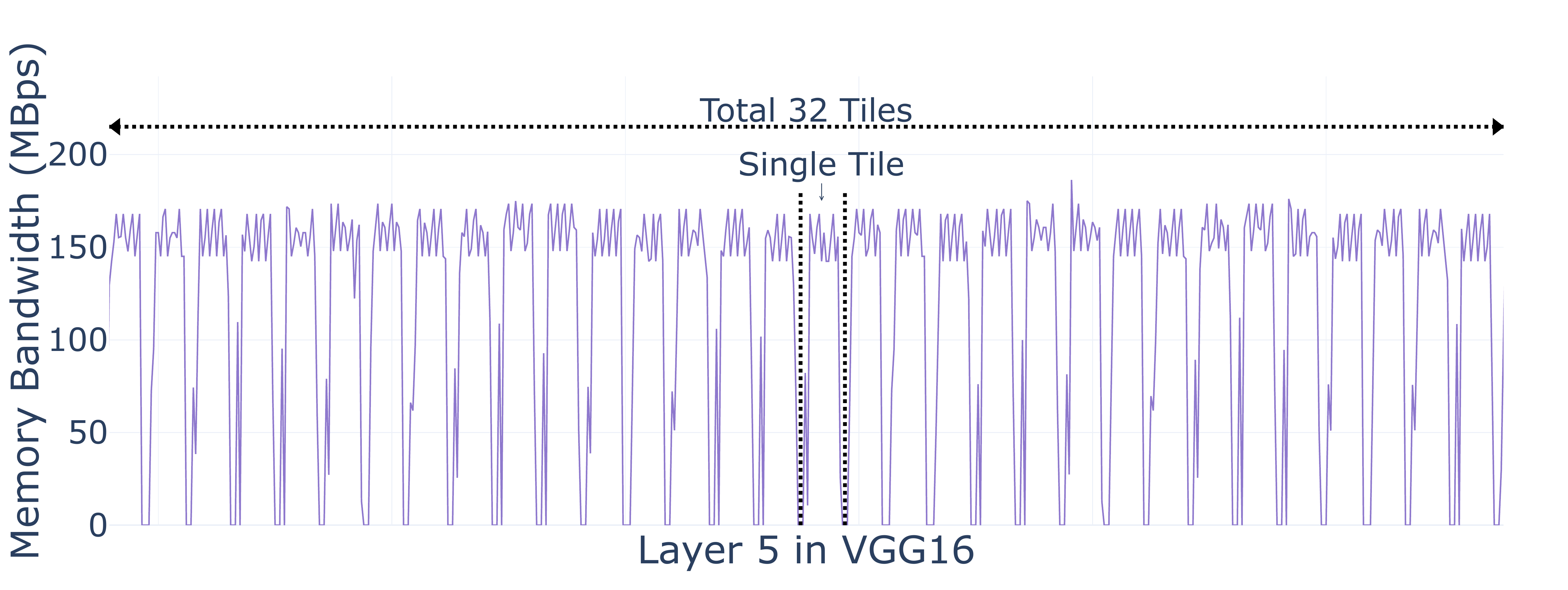}}
        \vspace{-0.5em}
        \subcaption{\label{fig:vgg16_layer5}\textbf{Read bandwidth of the 5th layer in vgg16.}}
    \end{subfigure}
    \caption{\label{fig:vgg16_attack_demo}\textbf{Different layers in vgg16 network utilizes different memory bandwidth, while bursts within each layer show number of tiles.}
    }
\end{figure}

\noindent\textbf{Memory bandwidth snooping:}
\label{subsec:mem_bw_attack}
Modern chips commonly include high precision memory bandwidth performance counters for performance debugging purposes. These may also be used by a cloud provider for ensuring quality of service(QoS) across multiple tenants using the same platform. However these performance counters can end up leaking side channel information. Figure~\ref{fig:vgg16_end2end} shows the memory traffic of all the convolution layers of VGG16. Each of the layers utilizes different amount of memory bandwidth based on tile size and number of tiles. 
Figure \ref{fig:vgg16_layer5} zooms in on layer 5 which consists of 32 tiles. The bandwidth trace leaks the number of tiles and the kernel size of the weight tensor. In Section~\ref{sec:evaluate_security}, we demonstrate that a classifier can detect all the boundaries based on change in traffic shape and bandwidth. Upon successful leakage of the structure of the model, the attacker can craft custom inputs and devise an attack to determine the weights\cite{scnn, ReverseCNN}.  Interestingly, this attack can be effected solely by observing bandwidth variations using performance counters even when the data and address buses are protected. To the best of our knowledge, our work is the first to make this observation and design
\system{} to defend against such \textit{observation channel} attacks (Section~\ref{subsec:traffic-shaper}).

\noindent\textbf{Shared dependency queues:}
The load, compute, and store units of the DAE accelerator are controlled through dependency queues that are shared among units.
An attacker can corrupt control dependencies of the victim program -- e.g., by triggering execution units before data loading -- and violate read-after-write (RAW) dependencies.
\system{} addresses this through dependency queue partitioning (Section~\ref{subsubsec:partitionDependencyQueue}).

\noindent\textbf{Shared access to scratchpads:}
Shared access to the scratchpad can help an attacker read out stale secrets belonging to the victim by computing on a scratchpad region without loading any data after a victim's execution.
\system{} uses partitioned scratchpads, private data zeroization logic, and instruction base/bound check
to address this threat
(Section~\ref{subsubsec:scratchpad}).

\noindent\textbf{Shared instruction queues:}
Shared instruction queues help the attacker perform cyclic attacks like prime-and-probe by inserting instructions between victim execution and creating a covert channel to leak data. \system{} design defends this threat with private instruction queues (Section~\ref{subsec:scheduler-instruction-queue}).

\noindent\textbf{Shared access to execution units:}
Shared instruction units can lead to cross-tenant contention attacks in a multi-tenant accelerator. \system{} schedules execution units among different tenants based on their program phases. 
\quad

\noindent\textbf{Variable time GEMM/ALU unit:}
Since ML weights are typically sparse, data driven optimizations leading to execution time variability leaking sensitive model/input data during spatial sharing. \system{} 
generates constant time execution instructions on private data operations (Section~\ref{subsec:constantEU}).


Vulnerabilities that require integrity protection (13) and those that employ the address bus side channel (14) assume the attacker to have physical access to the system. The \system{} prototype excludes these vulnerabilities since these are composable with our case-studies on confidentiality vs. on-chip attacks (by co-tenants and privileged software) and are not required in physically secured data-centers. Recent commercial products include encrypted DRAM without integrity protection \cite{AMD-SME,AWS-Graviton2,Intel-TME} for such use-cases. Nevertheless, solutions for integrity or address-buses \cite{MorphCount, Vault, Phantom, PathORM} can be composed with the \system{} prototype by extending the memory controller while our defenses address vulnerabilities that arise within the accelerator. 

\subsection{Sesame threat model variants}
\label{sec:ThreatModel}
 Our threat model is defined based on three parameters. \system defenses are configured based on settings of these three parameters, which are discussed below:
\begin{itemize}[leftmargin=*]
    \itemsep-0.25em
    \item \textbf{Multi-Tenant execution modes:} Cloud services deploy large accelerators capable of hosting multiple ML models simultaneously. We observe that there can be two modes of sharing accelerator resources across multiple tenants.  
    \begin{itemize}
        \itemsep0em
        \item \textbf{Temporal Sharing:} This corresponds to a scenario where a single tenant rents entire accelerator for a given duration. In this mode, the attack surface is restricted to observation/contention channels outside of the accelerator during job execution. After the execution is done attacker can read out secrets by extracting stale scratchpad data with use-after-free attacks.
        \item \textbf{Spatial Sharing:} In this mode multiple mutually distrusting model inference jobs are run concurrently in a single accelerator. The attacker can observe/contend for the memory bandwidth channel, shared scratchpad, dependency queues, shared buffer and execution units.
    \end{itemize}
\item \textbf{Private/secret model:} This parameter specifies whether model confidentiality is required and addresses scenarios where the ML service provider deploys his service on a third party cloud provider for economies of scale. 
\item \textbf{Private/secret user input:} This parameter specifies whether the of user input is confidential.
\end{itemize}

The threat model column in  Table~\ref{tab:VulAnal} identifies which vulnerabilities are relevant with respect to the later two of the three threat model parameters described above. It may be noted that vulnerability 8-12 apply to only spatial sharing mode while all others are applicable to both spatial and temporal sharing modes. Further in the threat model where both model and input are private the vulnerability list would be union of the individual lists.

Analog side channels like electromagnetic radiations and information leakage caused by accelerator power and temperature variations are orthogonal to our system.
Denial-of-Service (DoS) attacks by an attacker which either compromises the accelerator management software or tampers with the network connected between the CPU and the client or the accelerator and the CPU is also outside of our threat model.
We also do not protect against DoS attacks of a malicious kernel executed in the accelerator which leads to contention in shared resources.
The FPGA bitstream hash is checked during co-processor secure boot
and is assumed to be free of hardware trojans.
And runtime reconfiguration is disabled.

\section{SESAME Overview}
\label{sec:overview}

This section describes the overall architecture of \system{} to address new vulnerabilities (shaded blue in Table~\ref{tab:VulAnal}) that arise when an accelerator is shared by mutually distrusting tenants co-executing on untrusted infrastructure -- i.e., to build a \textit{multi-tenant accelerator enclave}. These vulnerabilities can be categorized under information leaks through observation-only channels (such as memory traffic or scratchpad access control) and contention-driven (e.g., on-chip queues, scratchpad and execution units) channels -- closing these with minimal slowdown requires both compiler and hardware support. In this section, we look at the overall secure accelerator architecture and highlight hardware structure that need shaping and partitioning. 

\begin{figure}[htbp]
\centerline{\includegraphics[width=\linewidth]{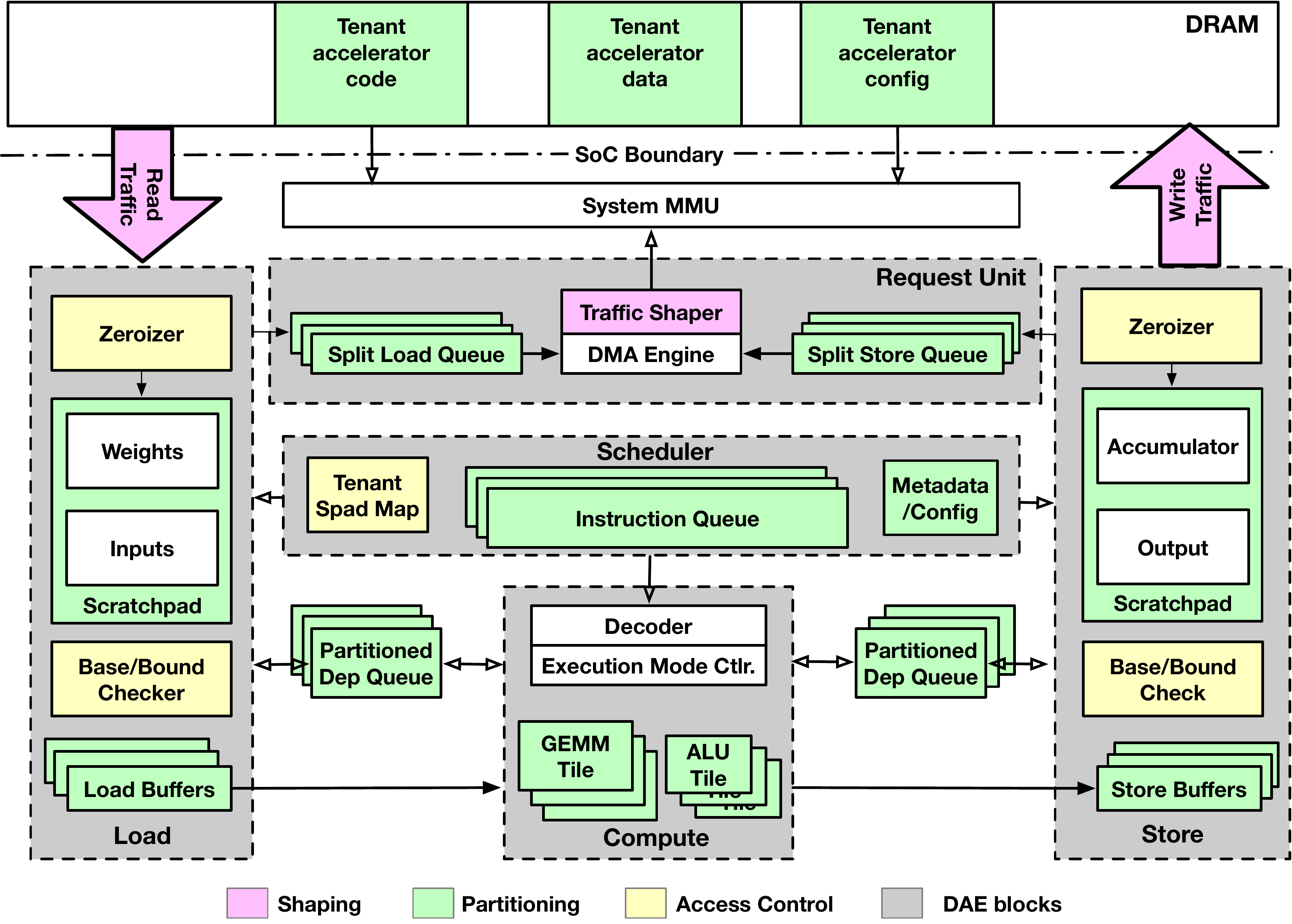}}
\caption{\textbf{\system{} design: a baseline DAE architecture extended with private queue partitioning, shaping, and access control blocks.}}
\label{fig:SystemDescUp}
\vspace{-1em}
\end{figure}

To bootstrap \system{}, a user authenticates the accelerator hardware and firmware using remote attestation protocols and a public-key certificate authority~\cite{Sanctum}. Users can compile their models on \system{} using host-side enclaves or remotely in a machine not exposed to attackers -- in both cases, a user transfers a \system{} binary to the accelerator and triggers execution until completion or for a fixed duration (if end-to-end timing channel is within the threat model).  
The binary along with the input data is loaded by the secure platform in DRAM and configuration registers specific to threat model requirements is programmed in accelerator registers.\\
\indent \system{} hardware comprises of three main primitives. The first component -- private queues -- secures shared queues that are used pervasively in hardware micro-architecture to decouple or pipeline function units. Private queues enable the accelerator software to partition shared queues across security domains and prevent information leaks through queue contention. 
%
The second component -- a traffic shaper -- closes leaks through all signals that come out an enclave (i.e., observation channels) by dynamically shaping the attacker-visible trace to look like a secret-independent distribution. 
The third component -- secure compilation passes -- takes as input programmer annotations to mark parts of the code and data as secret and generates accelerator instructions that minimize overheads from obfuscation-related data movement and cleanup (e.g., zeroing out scratchpads). In addition, \system{} includes additional logic to partition on-chip scratchpad memory, implement bound-checks for scratchpad accesses, and  hardware to clear out scratchpad space with dead private data. 
The key design principle is that private queues, traffic shapers, partitioned on-chip memory, and zeroization hardware can be composed arbitrarily. E.g., for small enclaves, shaping egress signals obviate the need to partition downstream queues, while large enclaves may be constructed by partitioning everywhere on-chip and using shapers exclusively for off-chip traffic.

Figure~\ref{fig:SystemDescUp} shows the components in \system{} hardware. The baseline DAE architecture includes a \textit{load unit} to load inputs and weights from memory into the scratchpad -- \system{} extends this with support to zero out memory and place bounds checks on scratchpad accesses to prevent buffer overflow. The \textit{store unit} similarly writes back outputs and intermediate values into memory and includes a scratchpad. The \textit{compute unit} performs the matrix arithmetic and GEMM computations that form the bulk of deep learning models -- \system{} requires both ALUs and GEMM units to support a constant-time mode where the execution time is independent of data value. The \textit{compute unit} is connected to both \textit{load} and \textit{store units} via dependency queues -- \system{} modifies these queues to be configurable as private queues. All traffic to memory is via a Request unit -- \system{} adds logic to shape the memory trace to this unit. The \textit{request unit} also has load, instruction, and store queues that a \system{} user can configure as private queues in spatial sharing mode.\\\\


\section{Programming Model:}

\begin{figure*}[htbp]
\centerline{\includegraphics[width=\linewidth]{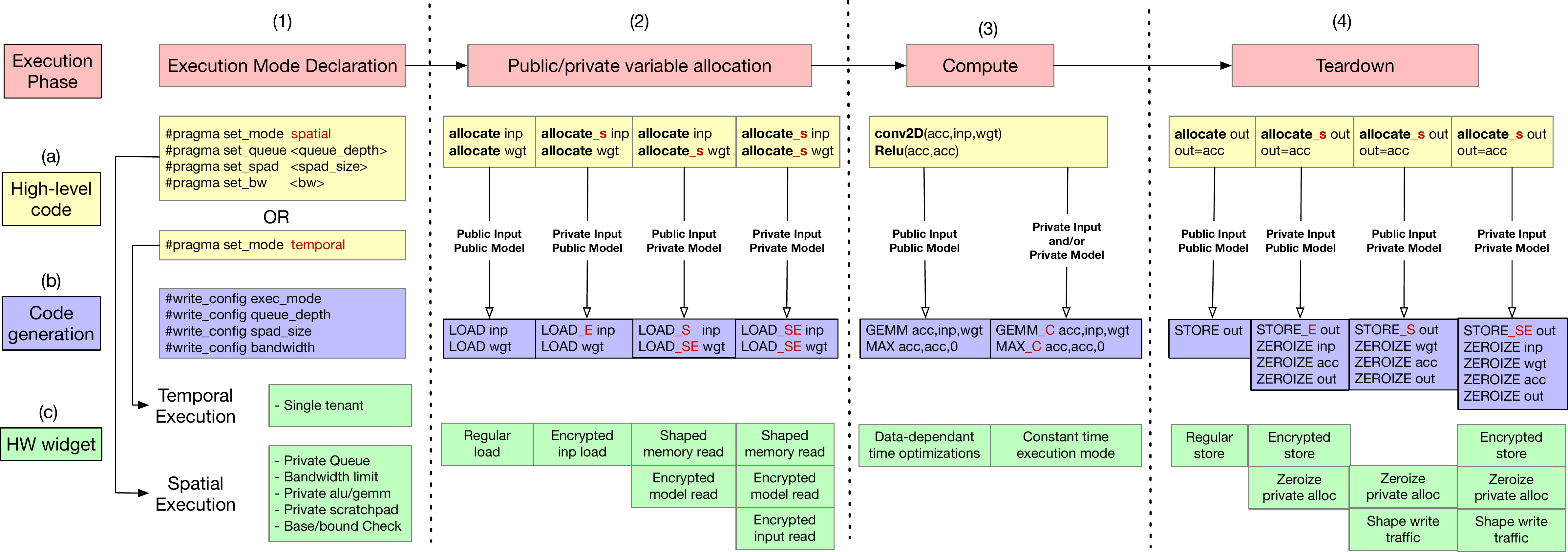}}
\caption{\textbf{Code generation example for various user-specified threat model invoking only the required secure hardware widgets}}
\label{fig:CodeGeneration}
\vspace{-1em}
\end{figure*}

\subsection{Instruction-set extensions:}
We introduce the following instructions to the baseline architecture:
\begin{itemize}[leftmargin=*]
    \itemsep-0.25em
    \item \texttt{LOAD\_E <spad\_range>,<dram\_range>}: Load secret data from DRAM that doesn't need traffic shaping but still needs to be decrypted. 
    \item \texttt{LOAD\_S(E) <spad\_range>,<dram\_range>}: Load data from DRAM with the traffic shaped read channel. Data decryption needed for the E variant.
    \item \texttt{GEMM\_C out,<inp1>,<inp2>}: This instruction disables all data-driven optimization of the gemm unit.
    \item \texttt{ALU\_C <out>,<inp>}: This instruction performs constant time ALU instructions. It prevents leaks through relu and clipping units. Input can be an immediate value as well.
    \item \texttt{ZEROIZE <spad\_range>}: This instruction is used to zeroize a portion of scratchpad address range. This instruction adds dependency to other instructions that uses conflicting regions of scratchpad. 
    \item \texttt{STORE\_E <dram\_range>,<spad\_range>}: Encrypt and store output from to DRAM.
    \item \texttt{STORE\_S(E) <dram\_range>,<spad\_range>}: Store data through a shaped memory write channel. Encryption needed by the E variant.
\end{itemize}

\subsection{High-level code Pragmas:}
\system{} supports application level annotation to identify secret data structures and specify execution mode. The software defined threat model as mentioned in section \ref{sec:ThreatModel} is propagated down to generate instructions and accelerator configurations through pragmas in high- level code. We describe below the various pragmas that are used to specify this information.
\begin{itemize}[leftmargin=*]
    \itemsep-0.25em
\item\textbf{allocate$\_$S}: This pragma, as illustrated in Figure~\ref{fig:CodeGeneration} is an enhanced version of the $allocate$ pragma used in the baseline which allocates scratchpad enables DMA transfer from DRAM. The $\_S$ annotation identifies secret data structures for the following purposes:
    \begin{itemize}[leftmargin=*]
        \itemsep-0.25em
        \item Data structures thus identified are stored in encrypted format in the memory and code generation unit appropriately annotates \texttt{load/store} instructions.
        \item It directs the code generation unit to generate \texttt{zeroize} instructions at scheduling boundaries to the scratchpad locations that hold these data structures and computation results generated from them.
        \item  Computational instructions operating on such structures are annotated to operate in constant time mode.
        \item Lastly, for threat scenarios where the traffic shaper has been enabled, this pragma helps annotate the \texttt{load/store} instructions that need the bandwidth defenses of the traffic shaper.  
    \end{itemize}
\item\textbf{exec$\_$mode}: It informs the driver if the application is sharing the accelerator with other tenants and need in-accelerator resource partitioning. 
\item\textbf{queue$\_$depth}: The driver requests the accelerator scheduler to reserve private queues for both instruction and dependency between the DAE components. This prevents contention attacks by an untrusted tenant co-executing in the accelerator in spatial sharing mode.
\item\textbf{spad$\_$size}: The driver programs the value specified by this pragma into tenant-private configuration space. The scheduler reserves regions of input, weight, accumulator and output scratchpad based on the size specified by this value. 

\item\textbf{bandwidth}: Memory-mapped registers corresponding to the traffic shaper are programmed with the constant bandwidth specified by this pragma for both read and write channels. Traffic generated by \texttt{LOAD\_S} or \texttt{LOAD\_SE} instructions is shaped to this bandwidth specification.
\end{itemize}

\subsection{Code transformations}
We envision the cloud provider providing entire accelerator(temporal sharing) or part of accelerator(spatial sharing) to each tenant. \system{} enables users to create enclaves that are tailored to their threat model.
Figure~\ref{fig:CodeGeneration} shows how a user specifies \textbf{(1)} the execution environment  to be temporally shared (single tenant at a time) or spatially shared (by multiple tenants simultaneously), and \textbf{(2)} whether their model and/or inputs are private(secret) or public.
\system{} then handles \textbf{(3)} accelerator compute execution and \textbf{(4)} tenant teardown, clearing out secrets after execution completes. Each of these stages starts with \textbf{(a)} high level user program description, \textbf{(b)} instruction or configuration metadata generation and \textbf{(c)} hardware widgets (e.g., queues and shapers) to isolate computation. Phase 1 creates the execution environment by enabling private queues and resource requirement. Phase 2 enables user to define private/public variables to determine the layer threat model, using which \system{} enables the hardware privacy widgets.
Phase 3 performs the constant-time computation on partitioned execution unit for private variables and regular execution for public operands. A tiled loop nest iterates over height,width and channel tiles. Convolution operation uses the GEMM units while the activation, pooling,batch normalization uses the ALU units. Finally, the result is stored back in phase 4 and the scratchpad state is zeroized to relieve on-chip resources of private data.

\section{Compiler Passes}
\label{sec:compilerPass}
\begin{figure}[hbtp]
\centering
\includegraphics[width=\linewidth]{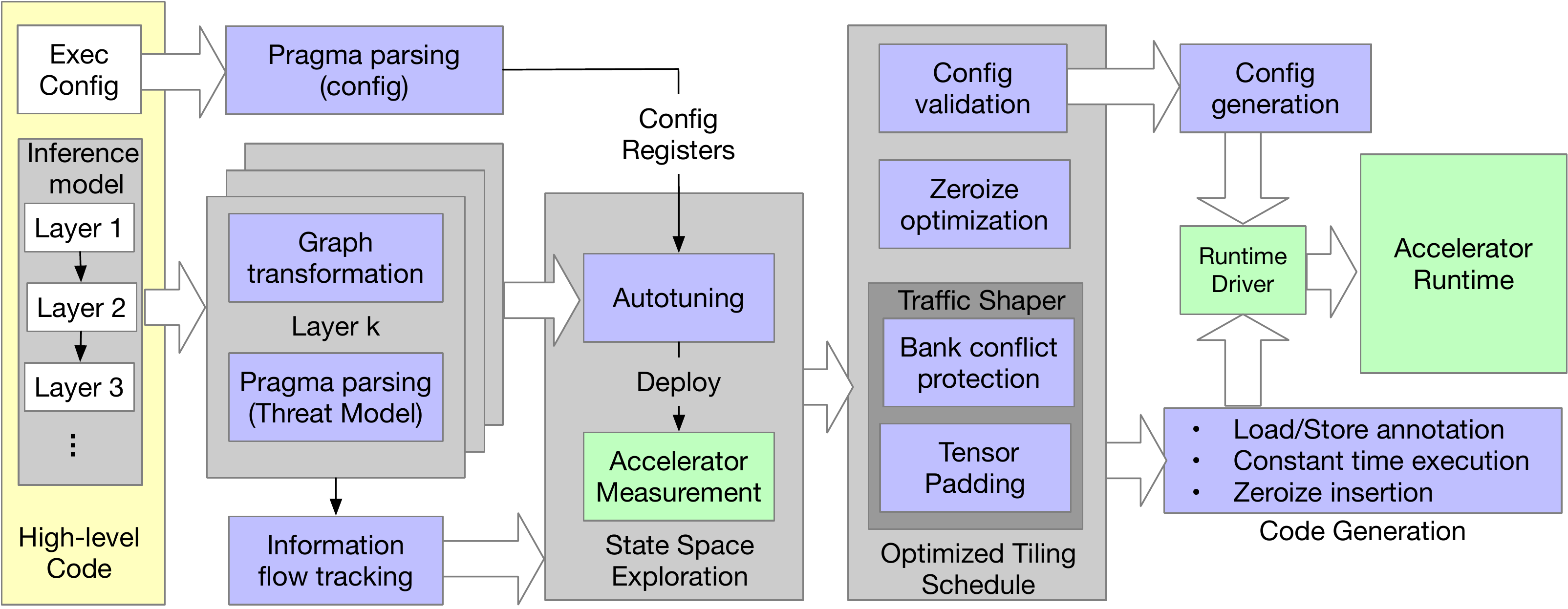}
\caption{\textbf{\system{} compiler passes play a key role in state-space exploration shaping and zeroization. Security passes are tightly coupled with user level threat model definitions}}
\label{fig:CompilerFlow}
\vspace{-1em}
\end{figure}

\system{} design encompasses enhancing the entire hardware/software stack for domain specific accelerators to support the different threat models described in Section~\ref{sec:ThreatModel}. This enables a user to get the best-possible performance within the threat model isolation realm. In this section we discuss how user specified security requirements filter through the various compiler passes and translate to setting up the required security constructs on the accelerator either through metadata for engaging hardware widgets or through instruction generation. Figure~\ref{fig:CompilerFlow} illustrates the various phases involved in this process which includes: \textbf{(1)} Perform graph transformation with TVM compiler on ML front and parse security pragmas and tag private variables information flow tracking for threat model specific transformations. \textbf{(2)} Auto-tuning phase perform tile-size exploration. \textbf{(3)} Traffic shaping and zeroize optimization on optimized tiling schedule. \textbf{(4)} Finally, the metadata after config validation and code generation phase ships the binary,data and config to the driver for deployment. 

We describe how specific compiler phases are adapted to the work on additional security information specified above:
\begin{itemize}[leftmargin=*]
    \itemsep-0.25em
    \item \textbf{Autotuning tiles:} 
    We use AutoTVM\cite{AutoTVM} framework to perform optimum tile search using threat model and resource constraints before accelerator deployment. Tile optimization chooses the best configuration for each layer for the entire model and ensures runtime tenant isolation.
    \item \textbf{Resource bill of materials (BOM):} Application resource requirement including memory bandwidth and execution mode is partially extracted from high-level code. The scratchpad, private queue and execution tile is extracted from the auto-tuning phase. This compiler pass eradicates illegal resource allocations limiting attackers to create resource contention and runtime errors.
    \item \textbf{Information flow tracking:} Any intermediate results generated from the sensitive data structures($allocate\_S$ variables) are marked as private. Scratchpad allocations of such variables are zeroized before de-allocation. This pass also flags warning to the user if a particular secret data is mistakenly marked as public at layer boundaries. Explicit dataflow tracking helps identify kernel schedules and variable liveness durations for precise allocation/zeroization.
    \item \textbf{Zeroize optimization:} \system{} compiler takes advantage of explicit kernel scheduling to reuse private scratchpad regions across different tiles without zeroizing it once at the layer boundary. This greatly reduces the binary size and scratchpad zero writes as shown in Figure~\ref{fig:ZeroOpt}.

    \item \textbf{Memory traffic shaper optimizations:} The memory traffic shaping is a software-hardware co-design. This pass is for protecting private ML models. \\
    \indent\textbf{Burst size equalization with padding:} The input and weight tensors are split into multiple equal-sized bursts. The tensor edges are padded to make it a multiple of burst size. Data is laid out such that no burst crosses a DRAM row-buffer boundary. The padding parameters are embedded in \system{} load instructions similar to VTA. Each burst is converted to AXI INCR transaction by the DMA engine.\\
    \indent\textbf{Memory bank conflict prevention:} Certain tiling configurations of input and weight tensors can cause bank conflicts which is reported in autotuning phase by the DMA engine. Since tiling on channel, height, and width axes is unique for each convolution layer, data layout transformation is done to ensure streaming data load on the inner loops and any bank conflict loads are rearranged to a different bank. 
    \item \textbf{Code generation:} Appropriate variants of \texttt{LOAD/STORE} instructions are generated for DMA transfer between memory and scratchpad. Compute instructions like \texttt{GEMM/ALU} instructions are annotated for constant operation to protect sensitive data structures. \texttt{ZEROIZE} instructions are generated to leave zero trace after private data execution.
    \item\textbf{Configuration generation:} Execution configuration is generated and tenant-private memory mapped registers are written to invoke threat-model specific isolation hardware widgets.

\end{itemize}

\section{Implementation}
\label{sec:implement}

Now, we discuss how the proof-of-concept implementation of memory traffic shaper, the partitioning of various queues, and the hardware scheduler is implemented to support multi-tenant enclaves. 

\subsection{Memory Traffic Shaper}
\label{subsec:traffic-shaper}
\begin{figure}[htbp]
\centering
{\includegraphics[width=\linewidth]{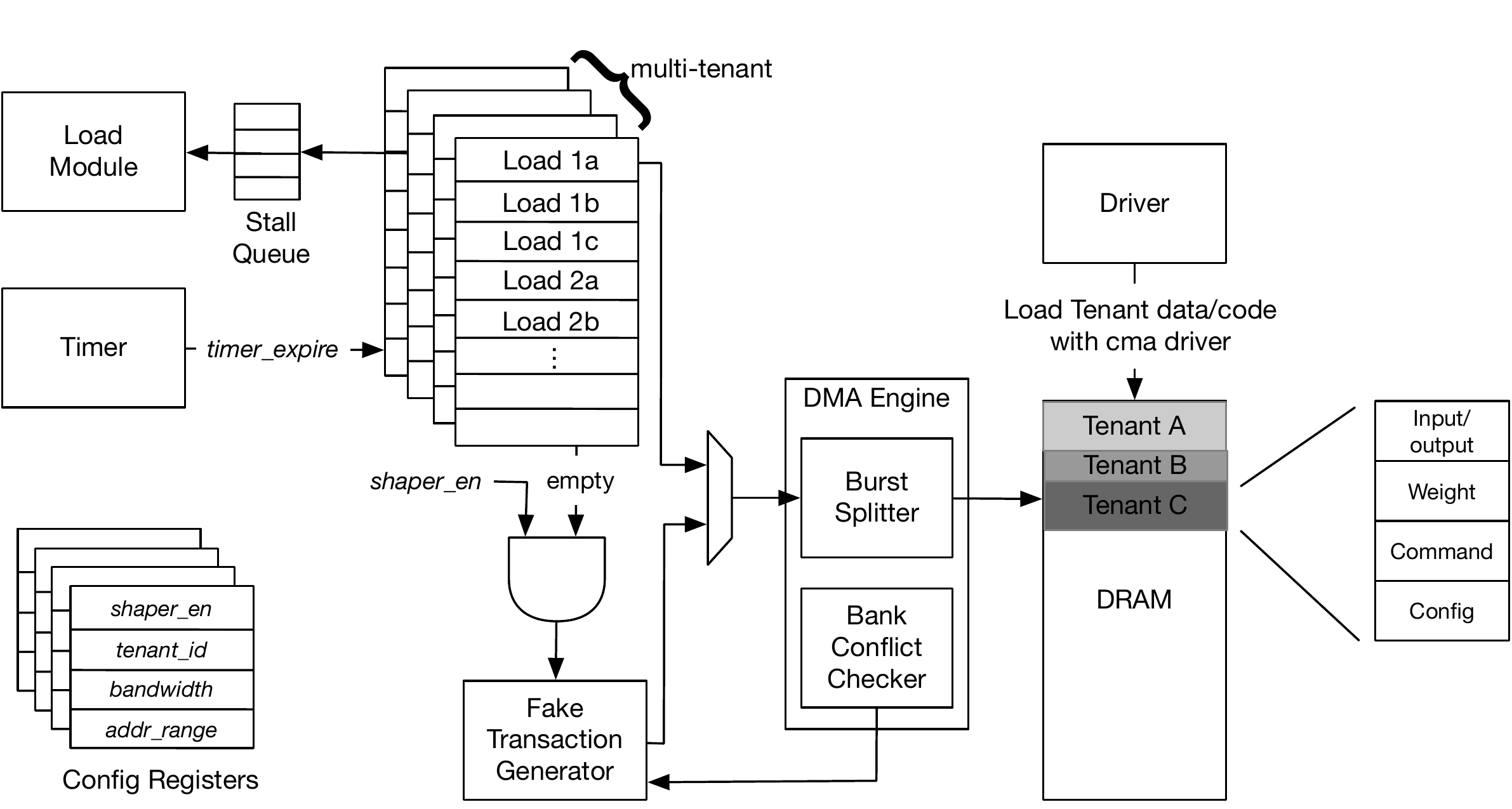}}
\caption{\textbf{Memory Traffic Shaper micro-architecture}}
\label{fig7_real}
\end{figure}

The memory traffic shaper ensures secret-independent data transfers from the DRAM to accelerator for a secret model application. It masks secret-dependent read/write bandwidth variations with a shaped trace with software programmed bandwidth during the lifetime of secret data computation and transfer. \\
\textbf{Compiler assumptions:} The \texttt{load/store} instructions are already a multiple of burst size and the data load by the driver ensures no bank conflict which are resolved in compiler as explained in section~\ref{sec:compilerPass}.\\
\textbf{Traffic shaper micro-architecture:}
Figure~\ref{fig7_real} shows the hardware components of the traffic shaper. A real transaction queue is filled with incoming real requests for each tenants. In order to provide model-size-independent trace, the dma engine always produces fixed-sized transaction bursts. A single load instruction is split into multiple equal sized bursts by the DMA engine and each burst is sent on $timer\_expiry$ determining the overall bandwidth. Moreover, to hide the computation and data dependency at runtime, the traffic generator has a fake transaction generator to produce transactions to free memory banks. A real transaction fifo full signal prevents the load module from generating further memory requests. Same hardware logic is present for write channel.\par
The traffic shaper has tenant private configuration registers. \textit{Shaper\_en} of each tenant(identified by \textit{tenant\_id}) enables the fake transaction generator and limits the transaction generator to produce constant sized transactions. The \textit{bandwidth} register programs the timer for each tenant to ensure bandwidth QoS and the \textit{addr\_range} enables the fake transaction generator to generate requests within the memory-mapped address range to prevent access-control violations.\\

\subsection{Partitioning Resources}
\label{subsec:partition-resources}
In-accelerator partitioning for the following resources is done for spatial multi-tenancy to deter on-chip attacks:

\subsubsection{Dependency queues}
\label{subsubsec:partitionDependencyQueue}
The DAE dependency queues are partitioned four ways in our implementation to enable multiple tenants. This split prevents a tenant to corrupt/contend the control dependency with a co-executing tenant. Tenant ID is used to redirect queue read/writes to the tenant-private one. The queue depth comes from configuration register and is validated during resource BOM compiler phase. 

\subsubsection{Scratchpad}
\label{subsubsec:scratchpad}
\system{} has four scratchpads -- Weight, Accumulator, Input, and Output -- with capacities listed in Table~\ref{tbl:SystemSpecifications}. Each scratchpad is split into $16kB$ regions whose tenant ownership is maintained in a scheduler \textit{scratchmap} data structure. Scratchpad address base and bound logic checks tenant ownership from \textit{scratchmap} before each access keeping tenants isolated. Zeroizer logic clears private data and tenant ownership at scratchpad region granularity during teardown.

\subsubsection{Execution units} The \textit{GEMM} and \textit{ALU} execution units are spatially partitioned into four 8x8 units with private load/store buffers to eradicate execution unit contention in the multi-tenant scenario. This utility is useful for smaller kernels co-executing in large accelerators like TPU\cite{TPU} and cloud accelerator deployment of \system{}. 

\subsection{Scheduler}
\label{subsec:scheduler-instruction-queue}

\system{} has a hardware scheduler block consisting of the instruction queue and data structures keeping track of each tenant's resource occupancy and guarantee runtime computational non-interference. It houses tenant maps for dependency queues, scratchpad, execution tile and instruction queue providing complete on-chip separation between multiple tenants. It also communicates with the driver to launch new tenant and sets configuration register notifying application teardown.

\subsection{Constant time execution units}
\label{subsec:constantEU}
\system{} ISA supports constant-time execution instructions (e.g. \texttt{gemm\_c},\texttt{alu\_c}) but our PoC implementation only includes 8-bit DSP implementation of execution units.\revision{ The performance headroom for such optimizations is limited for quantized ML inference.}
\subsection{Private data encryption}
\label{subsec:memEnc}
\revision{To estimate the performance overhead of encryption, \system{} PoC implementation emulates QARMA128 with a $10ns$ delay and AES128\cite{AES} with a $20ns$ delay for every 128-bit private data access. These delay numbers are taken from prior work\cite{RWC_Avanzi}.}
\section{Evaluation}
\system{} proof-of-concept(PoC) implementation is built by enhancing the VTA\cite{VTA} hardware baseline with a constant traffic shaper to mask the memory bus traffic, partitioned on-chip accelerator resources like dependency queues and scratchpads for tenant isolation, a zeroization module to clear private data, an address base/bound logic for access checks and a hardware scheduler for runtime tenant isolation management. The prototype runs on a Xilinx Pynq-Z1 board with a dual-core arm CortexA9 co-processor and an accelerator prototype in FPGA fabric with specifications listed in table \ref{tbl:SystemSpecifications}.\par

\begin{table}[ht]
    \centering
    \begin{tabular}{|c|c|c|}
    \hline
    \RCG{20} 
    \multicolumn{1}{c|}{\color{white}\textbf{Component}} & \multicolumn{2}{c|}{ \color{white}\textbf{Specification}} \\ [0.5ex] 
    \cline{1-3}
    \multicolumn{1}{|p{3cm}|}{Processor} & \multicolumn{2}{p{4.5cm}|}{Dual arm CortexA9 @667MHz}\\ 
    \multicolumn{1}{|p{3cm}|}{DRAM} & \multicolumn{2}{p{4.5cm}|}{512MB DDR3 @ 533MHz} \\
    \multicolumn{1}{|p{3cm}|}{FPGA frequency} & \multicolumn{2}{p{4.5cm}|}{Zynq 7020 @ 100MHz} \\
    \hline \hline
    \multicolumn{1}{|c|}{\textbf{\textit{Accelerator}}} & \multicolumn{1}{|c|}{\textbf{\textit{temporal}}} & \multicolumn{1}{|c|}{\textbf{\textit{spatial}}}\\
    \hline \hline
    \multicolumn{1}{|p{3cm}|}{Weight Scratchpad} & \multicolumn{1}{|p{2.25cm}|}{2MB} & \multicolumn{1}{|p{2.25cm}|}{512KB}\\
    \multicolumn{1}{|p{3cm}|}{Input Scratchpad} & \multicolumn{1}{|p{2.25cm}|}{256KB} & \multicolumn{1}{|p{2.25cm}|}{64KB}\\
    \multicolumn{1}{|p{3cm}|}{Output Scratchpad} & \multicolumn{1}{|p{2.25cm}|}{256KB} & \multicolumn{1}{|p{2.25cm}|}{64KB}\\
    \multicolumn{1}{|p{3cm}|}{Acc Scratchpad} & \multicolumn{1}{|p{2.25cm}|}{512KB} & \multicolumn{1}{|p{2.25cm}|}{128KB}\\ 
    \multicolumn{1}{|p{3cm}|}{GEMM units} & \multicolumn{1}{|p{2.25cm}|}{256} & \multicolumn{1}{|p{2.25cm}|}{64}\\
    \multicolumn{1}{|p{3cm}|}{Memory bandwidth} & \multicolumn{1}{|p{2.25cm}|}{400MB/s} & \multicolumn{1}{|p{2.25cm}|}{100MB/s}\\
    \hline    
    \end{tabular}
    \caption{\revision{\textbf{System Specifications with per-tenant temporal and spatial sharing resource allocation}}} 
    \label{tbl:SystemSpecifications}
    \vspace{-1em}
\end{table}

To simplify the implementation, the resource BOM which includes traffic shaper bandwidth, datasize of each layer etc. is extracted manually after compiler autotuning phase and fed to the runtime driver as command line parameters. The driver bootstraps the accelerator by creating a config,code and data memory-mapped address space as shown in Figure~\ref{fig:SystemDescUp}. The config address space houses live resource availability status and tenant-private metadata regions for configuration loading. The proof-of-concept implementation supports upto four tenants.
The accelerator driver begins by querying resource availability registers and schedules a new tenant if adequate resources are available to prevent over-subscription and creates a tenant ID. The driver then loads instructions into the tenant-private instruction region and input/model values into the data region. After that the driver waits for the results by polling a configuration register.

In this section, we evaluate the hardware prototype by running \textit{imagenet} inference of six trained deep learning networks taken from MXNET modelzoo\cite{MxNet}.
The models are quantized to support 8-bit operation.
The image classification models chosen are VGG11, VGG16, AlexNet, ResNet18, ResNet34, and ResNet50.
We first assess the security provided by traffic shaper against memory bandwidth side channel attacks, followed by instruction binary size and finally a performance comparison for all threat models for spatial and temporal execution modes. The four bars of each plot is for \textbf{(1)} Public Input Public Model; \textbf{(2)} Private Input Public Model;\textbf{(3)} Public Input Private Model; \textbf{(4)} Private Input Private Model.

\subsection{Security Evaluation}
\label{sec:evaluate_security}
\subsubsection{Memory traffic shaper}
In this section we analyze the effectiveness of \system{} with the traffic shaping primitive discussed in Section~\ref{subsec:traffic-shaper} to 
protect against a memory bandwidth side channel attack from section~\ref{subsec:mem_bw_attack}.
To show that bandwidth variations is a problem and validate the traffic shaper, a bandwidth measurement widget is synthesized along with the FPGA bitstream. The bandwidth measurement widget counts the AXI read and write channel data bytes transferred to report memory bandwidth.   
Figure~\ref{fig:MemoryTrace} shows the memory traffic pattern before and after shaping collected for the six workloads. Both read/write bandwidth in unshaped trace is leaky and provides kernel size and layer boundary information. The figure shows a single run but each benchmark is run fifty times and the median is constant for every sample. Nevertheless, a two stage classifier is designed to attack both the unshaped and the shaped trace as follows:

\begin{figure}[t!]
    \centering
    \includegraphics[width=\linewidth]{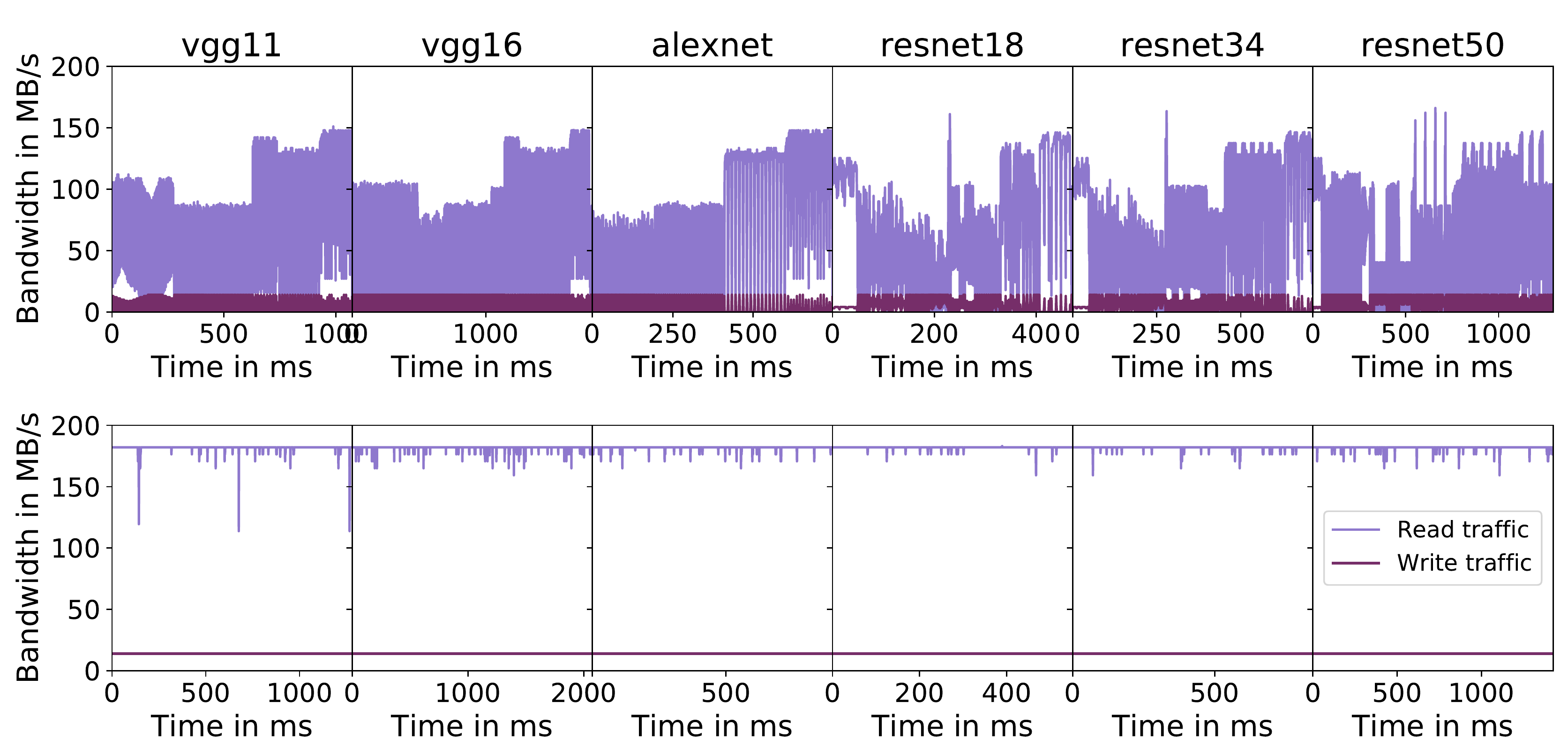}
    \caption{\textbf{Comparison between real and shaped memory traffic.
    Top figure shows the real read/write traffic of each network
    while the bottom one plots the shaped traffic.}}
    \label{fig:MemoryTrace}
    \vspace{-1em}
\end{figure}

\newcommand{\ra}[1]{\renewcommand{\arraystretch}{#1}}

\begin{table}[t]
\resizebox{1.01\linewidth}{!}{%
\ra{1.1}
\begin{tabular}{cccccccccccccc}
\hline
\multirow{3}{*}{} & &
\multicolumn{2}{c}{AlexNet}  & \multicolumn{2}{c}{VGG11} & 
\multicolumn{2}{c}{VGG16}    & \multicolumn{2}{c}{ResNet18} & 
\multicolumn{2}{c}{ResNet34} & \multicolumn{2}{c}{ResNet50} \\ 
\cline{3-14} 
& &
\begin{tabular}[c]{@{}c@{}}easy\\ 3\end{tabular} & 
\begin{tabular}[c]{@{}c@{}}all \\ 4\end{tabular} & 
\begin{tabular}[c]{@{}c@{}}easy\\ 5\end{tabular} & 
\begin{tabular}[c]{@{}c@{}}all \\ 6\end{tabular} & 
\begin{tabular}[c]{@{}c@{}}easy\\ 8\end{tabular} & 
\begin{tabular}[c]{@{}c@{}}all \\ 11\end{tabular} & 
\begin{tabular}[c]{@{}c@{}}easy\\ 22\end{tabular} & 
\begin{tabular}[c]{@{}c@{}}all \\ 23\end{tabular} & 
\begin{tabular}[c]{@{}c@{}}easy\\ 24\end{tabular} & 
\begin{tabular}[c]{@{}c@{}}all \\ 36\end{tabular} & 
\begin{tabular}[c]{@{}c@{}}easy\\ 50\end{tabular} & 
\begin{tabular}[c]{@{}c@{}}all \\ 52\end{tabular} \\
\hline
\hline
& \begin{tabular}[c]{@{}c@{}}precision\end{tabular} &
1 & 1 & 1 & 1 & 1 & 1 & 0.69 & 0.64 & 0.66 & 0.72 &  0.33 & 0.33 \\ \\
\multirow{-3}{*}{\rotatebox[origin=c]{90}{\footnotesize{Unshaped}}} & 
\begin{tabular}[c]{@{}c@{}}recall\end{tabular} &
0.75 & 1 & 0.83 & 1 & 0.73 & 1 & 0.96 & 1 & 0.67  & 1 & 0.96 & 1\\ \hline
& \begin{tabular}[c]{@{}c@{}}precision\end{tabular} &
0.03 & 0.03 & 0.01 & 0.01 & 0.0027 & 0.00011 & NA & NA & NA & NA & NA & NA \\  \\
\multirow{-3}{*}{\rotatebox[origin=c]{90}{\footnotesize{Shaped}}} & 
\begin{tabular}[c]{@{}c@{}}recall\end{tabular} &
0.75 & 1 & 0.83 & 1 & 0.73 & 1 & NA & NA & NA & NA & NA & NA \\ \hline
\end{tabular}%
}
\caption{\textbf{Precision and recall when identifying layer boundaries for each network.
The second row in this table lists the number of the boundaries between two consecutive layers that are of different types (easy) and the total number of layer boundaries (all).
The precision and recall are calculated when the attacker detects all boundaries while introducing false positives (if any).
NA indicates that the attacker fails to identify the boundaries while introducing over $10000\times$ false positives.}}
\label{tbl:membw-channel-fpres}
\end{table}

\newcommand{\mnrow}[2][c]{%
      \begin{tabular}[#1]{@{}c@{}}#2\end{tabular}}

\begin{table}[t]
\centering
\resizebox{\linewidth}{!}{%
\begin{tabular}{cccccccc}\toprule
    & AlexNet & VGG11 & VGG16 & ResNet18 & ResNet34 & ResNet50 & Overall \\
\midrule
\mnrow{Execution \\time only}&
1 & 1 & 0.958  & 0.896 & 0.851 & 0.824 & 0.826 \\
\midrule
\mnrow{SVM w/ \\(w/o) DWT} &
1 & 1 & 1 & 1     & 1 & \mnrow{0.811} & \mnrow{0.927} \\ 
\midrule
\mnrow{MLP w/ \\(w/o) DWT} &
1 & 1 & 1 & 1     & \mnrow{1\\(0.986)} & \mnrow{0.868\\(0.849)} & \mnrow{0.949\\(0.934)} \\  
\midrule
\mnrow{CNN w/ \\(w/o) DWT} &
1 & 1 & 1 & 1     & 1 & \mnrow{0.877\\(0.830)} & \mnrow{0.953\\(0.934)}\\ 
\bottomrule
\end{tabular}%
}\caption{\label{tbl:layerClassificationAccuracy}%
\textbf{Layer type classification accuracy using unshaped traffic assuming perfect layer boundary detection.
The first row shows the accuracy only using the termination timing of the layers for classification.
\textit{It serves as a baseline accuracy as long as attacker can identify the layer boundaries.}
2-4th rows show accuracy of 3 classifiers using bandwidth trace
with and without frequency domain signal computed from discrete wavelet transform (DWT).}
}\end{table}

\indent\textbf{Layer Boundary Detector:}
Prior arts~\cite{ReverseCNN,DeepSniffer} demonstrated that
the read-after-write (RAW) pattern on the
address trace reveals the layer boundaries accurately but bandwidth variation on read/write channel is used to detect layer boundaries.
We first use the RAW dependency pattern to identify
potential boundaries (the end of each write transaction) on unshaped trace. We then compare the memory bandwidth within a fixed time window before and after the potential boundary.
Statistics like total read/write data, median and peak bandwidth, standard deviation as well as frequency domain signals computed using discrete wavelet transform (DWT) are used.
For shaped trace,
since write bus is constantly exercised, the RAW activity is invisible to the attacker.
Instead, we model an attacker who offline profiles termination timing of
all possible layer configurations with termination timing protection turned-off.
At run-time, the attacker use a combination of
termination timings to enumerate potential layer boundary candidates.
This helps constrain the maximum false positives.
Similarly, at a boundary candidate, attacker compares the memory bandwidth channel before and after the candidate using the same statistics.
The different thresholds for this determination
allows the attacker to trade-off the aggressiveness in boundary identification vs.~triggering false positives.

Table~\ref{tbl:membw-channel-fpres} shows the precision and recall numbers for both unshaped and shaped memory traffic
on our evaluation model suite.
The second row in the table shows the total number of layer boundaries
as well as the number of easily identifiable boundaries (adjacent
layers utilize different memory bandwidth) from the unshaped traffic.
Without the traffic shaper in place, the classifier can detect easy boundaries with 100\% precision for AlexNet, VGGs and ResNet18.
Note that for ResNet models, the residual layers are very short,
which makes the boundaries in between them hard to detect even without traffic shaper.
With the traffic shaper in place, the precision drops down to as low as 0.01\% for VGG16.
NA in the table indicates that our modeled attacker fails to identify every layer boundary
in three ResNet models,
while triggering over $10000\times$ false positives (precision $<0.0001$).

\indent\textbf{Layer Type Classification.}
Compared to one feature vector per kernel in Deepsniffer~\cite{DeepSniffer},
fine-grained observations enable attackers to model each layer
as a time-series of bandwidth information.
In addition to memory bandwidth related features used in DeepSniffer,
we include frequency domain signals(DWT) to capture IFM tile load/store memory bandwidth signatures.
Our training data constitutes of bandwidth traces from different potential layer configurations.
we test victim memory traffic time-series using three classifiers: Support Vector Machine (SVM),
Multilayer Perceptron (MLP), and Convolutional Neural Network (CNN),
with and without frequency domain signals. 
Similar to~\cite{DeepSniffer}, SVM and MLP use one feature vector per layer, while CNN uses a sequence of feature vectors for classification, enabled by fine-grained observation.
Table~\ref{tbl:layerClassificationAccuracy} shows the layer-type classification accuracy with unshaped traffic data after the layer boundaries are identified.
The first row shows classification accuracy merely using the execution length of each layer.
\textit{This is a baseline accuracy for any attacker with the knowledge of the layer boundaries (execution timing of layers).}
The 2-4th rows show accuracy of the three evaluated layer-type classifiers.
From execution-time based classifier to bandwidth-based classifiers,
the accuracy jumps from 84\% to 93\% on average.
From SVM to CNN, accuracy increases with increasing classifier complexity.
In addition, including frequency domain signals improves the accuracy
as different tile size configurations result in different compute/IO ratio.

However, after applying traffic shaping, the classifiers
are not able to classify features among different layer types.
The resulting accuracy
is similar to the baseline attacker knowing only the layer termination times.
But as the layer boundaries for the shaped trace is undetectable (Table~\ref{tbl:membw-channel-fpres}),
we conclude that \system{} seals
the memory read/write bandwidth channel.

\subsubsection{Partitioned logic}
The partitioning and access control hardware is validated first at RTL level by system verilog tests. Runtime validation is done by changing a tenant binary in after code generation to access an unauthorized scratchpad location. The access was blocked on both read/write channel in \textit{BRAM}.Moreover, scratchpad read after tenant teardown returned zero value.

\subsection{Compiler Evaluation}
\subsubsection{Tile optimization performance} 
\begin{figure}[htbp]
    \centering
    \includegraphics[width=\linewidth]{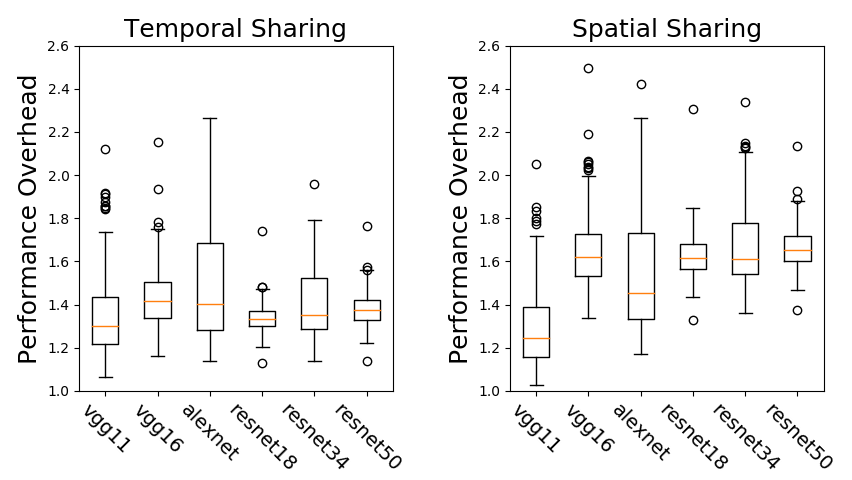}
    \caption{\textbf{Performance overhead variation for design space exploration of tile size configuration}}
    \label{fig:TileOpt}
    \vspace{-1em}
\end{figure}

 Figure~\ref{fig:TileOpt} illustrates the variability of performance overhead for the threat model where both input and model are private.  All of the accelerator resource is used for temporal sharing and 4 tenants equally share accelerator resource in spatial sharing. Each benchmark is run with 800 different tile combinations across layers with overhead comparison with baseline VTA. Tile optimization helps maximum available resource utilization and  the best tile-configurations change with resource bom. It results in better utilization of the traffic shaper bandwidth for private model.

\subsubsection{Zeroize Optimization}
\begin{figure}[htbp]
    \centering
    \includegraphics[width=\linewidth]{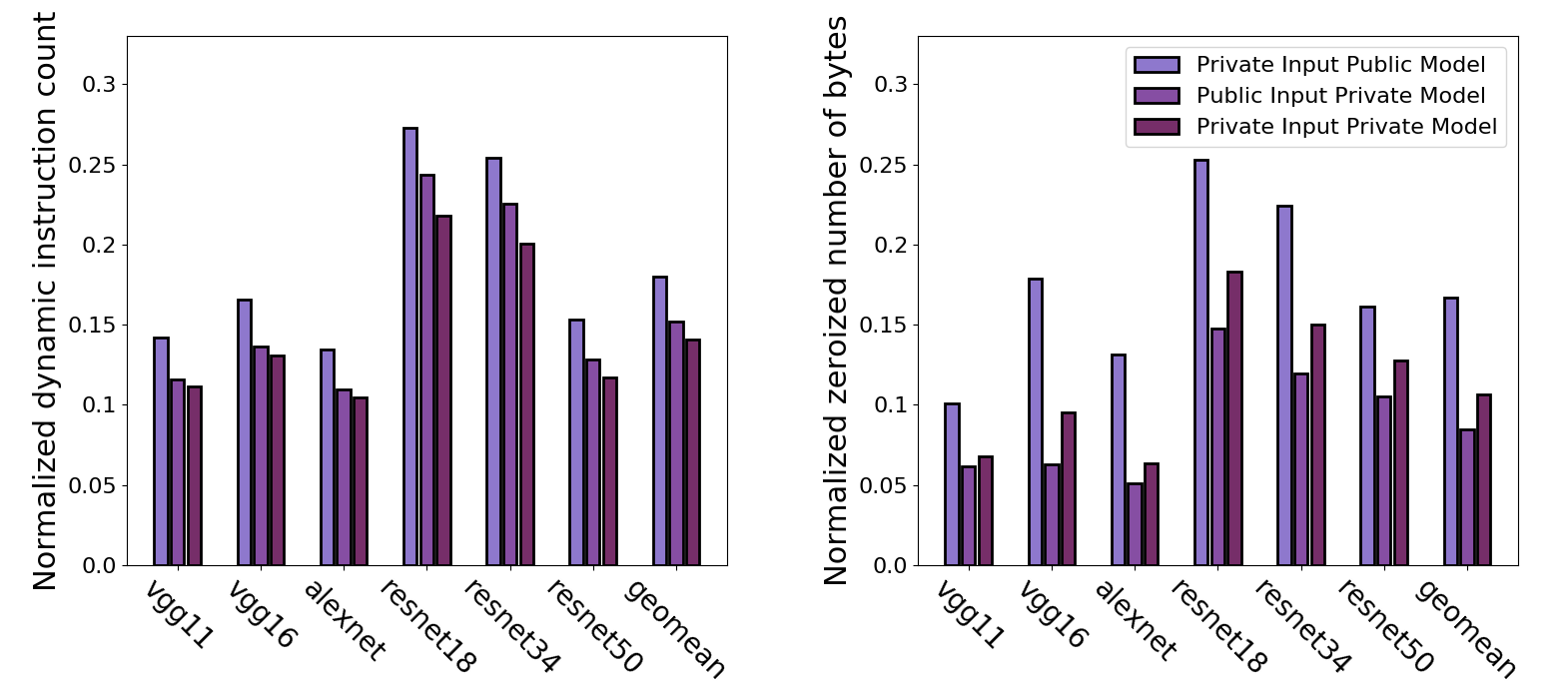}
    \caption{\textbf{Dynamic instruction count and number of zeroized BRAM bytes normalized wrt zeroizing every private data for different threat models}}
    \label{fig:ZeroOpt}
    \vspace{-1em}
\end{figure}

Figure~\ref{fig:ZeroOpt} shows the reduction in the size of scratchpad regions that need to be zeroized over a naive case of zeroizing all private data held by scratchpad. Kernel scheduling helps re-utilize private scratchpad regions across multiple tiles that share the same security level. Only zeroizing regions before loading public data reduces dynamic zeroize instruction count by 14\% to 18\% based on the threat model. Since each instruction clears variable sized BRAM regions, the right plot shows the number of BRAM bytes zeroized. There is a reduction between 8.5\% and 16.7\% across different threat models. There is higher reduction in weight secret threat model for vgg11,vgg16 and alexnet due to larger kernel sizes as compared to the resnets. The private model(bar 2) shows higher reduction than private input(bar 1) and the private model and secret threat model(bar 3) plot is closer to the weight reduction percentage due to higher number of channels in model weight than the IFM/OFM. 

\subsubsection{Code Generation Instruction Mix}
\begin{figure}[hbtp]
\centerline
{\includegraphics[width=1.05\linewidth]{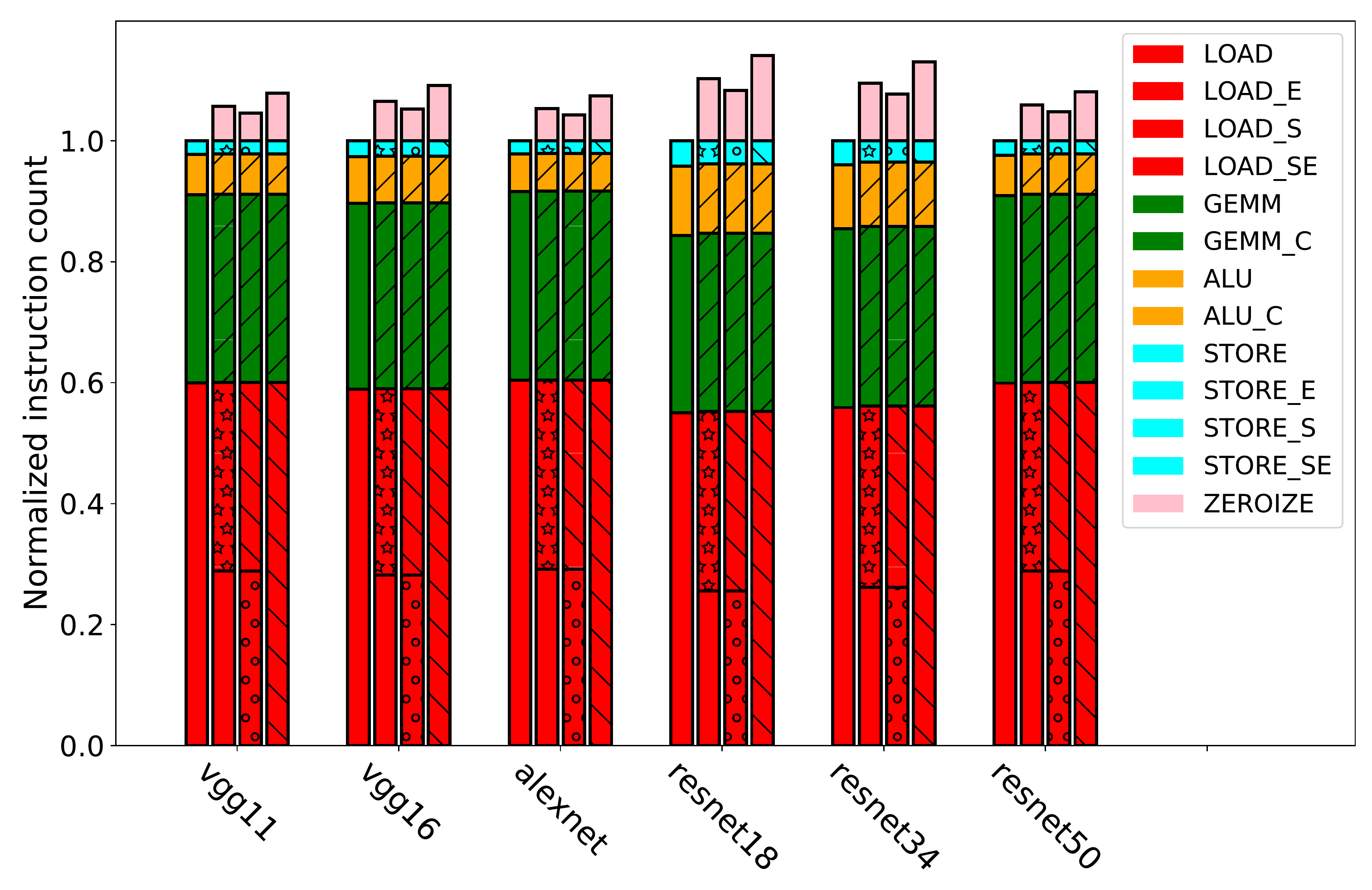}}
\caption{\textbf{Instruction mix for different threat models}}
\label{fig:instmix}
\vspace{-1em}
\end{figure}

Figure \ref{fig:instmix} shows the normalized instruction mix for six benchmarks for different threat models. Increase in binary size is solely due to addition of \texttt{zeroize} instructions. Load/store instruction variants change with threat model. \texttt{Load\_e} instructions are generated for loading private input when model is public. while a private model needs a read/write traffic shaper and uses \texttt{load\_se}/\texttt{store\_se} instructions. When the model is private, public input may be stored unencrypted in the DRAM and uses \texttt{load\_s} as its traffic needs to be shaped to hide access patterns. When both model and input are private all read accesses are performed using the \texttt{load\_se} instructions. Internal scratchpad locations holding secret data are cleared with \texttt{zeroize} instructions which varies based on the threat model. Each \texttt{zeroize} instruction clears different amount of scratchpad regions. Even though the number of \texttt{zeroize} instruction is higher in private input, the latency is higher in private model as \texttt{zeroize} clears a larger model scratchpad region.

\subsection{Performance/FPGA Utilization Results}

\subsubsection{FPGA Utilization Overhead}

Table \ref{tab:util} lists the percentage of FPGA resource used for each design component. The scheduler uses distributed RAMs to store tenant ownership of scratchpad partitions and private queues. The additional configuration registers lead to increase in register and LUT logic. Memory traffic shaper uses BRAM for tenant-private load/store queues. Distributed ram is used in DMA engine to partition loads and storing pending transaction bank information. Glue logic and timer accounts for the increase in register/LUT. The same queue size in baseline is split into multiple sections and access control glue logic accounts for area increase in partitioned resources.

\begin{table}[htbp]
    \centering
    \resizebox{1\columnwidth}{!}{%
    \begin{tabular}{|c|c|c|c|c|c|}
\hline
\RCG{20}
 \WT{20}Component & \WT{20}Baseline & \WT{20}Scheduler & \WT{20}\mrow{Memory\\Traffic\\Shaper} & \WT{20}\mrow{Partitioned\\Resources} & \WT{20}\mrow{SESAME\\Prototype} \\ \hline \hline
\mrow{Logic\\LUT}  & \RC 46.97\% & \mrow{9.57\%\\(1.2x)}& \mrow{2.56\%\\(1.05x)}& \mrow{2.4\%\\(1.05x)}& \RC \mrow{61.5\%\\(1.3x)}\\ \hline 
Register & \RC 19.5\% & \mrow{11.26\%\\(1.6x)}& \mrow{6.24\%\\(1.3x)}& \mrow{3.6\%\\(1.2x)}& \RC \mrow{40.76\%\\(2.1x)} \\ \hline
BRAM & \RC 92.14\% & \mrow{1.6\%\\(1.02x)}& \mrow{3.22\%\\(1.04x)}& \mrow{0\%\\(0)}& \RC \mrow{96.96\%\\(1.05x)}\\ \hline
\mrow{Distributed\\RAM} & \RC 11.51\% & \mrow{10.27\%\\(1.9x)}& \mrow{3.3\%\\(1.29x)}&  \mrow{0\%\\(0)}& \RC \mrow{23.08\%\\(2.178x)}\\ \hline
DSP & \RC 100\% & \mrow{0\%\\(1x)} & \mrow{0\%\\(1x)}& \mrow{0\%\\(1x)}& \RC \mrow{100\%\\(1x)} \\ \hline
    \end{tabular}
    }
    \caption{\textbf{FPGA Component Utilization}}
    \label{tab:util}
    \vspace{-1em}
\end{table}

\vspace{0.1in}
\subsubsection{Performance Overhead}
\begin{figure}[t]
    \centering
    \begin{subfigure}{1.05\linewidth}
        \centering
        \hspace{-1em}
        \includegraphics[width=\linewidth]{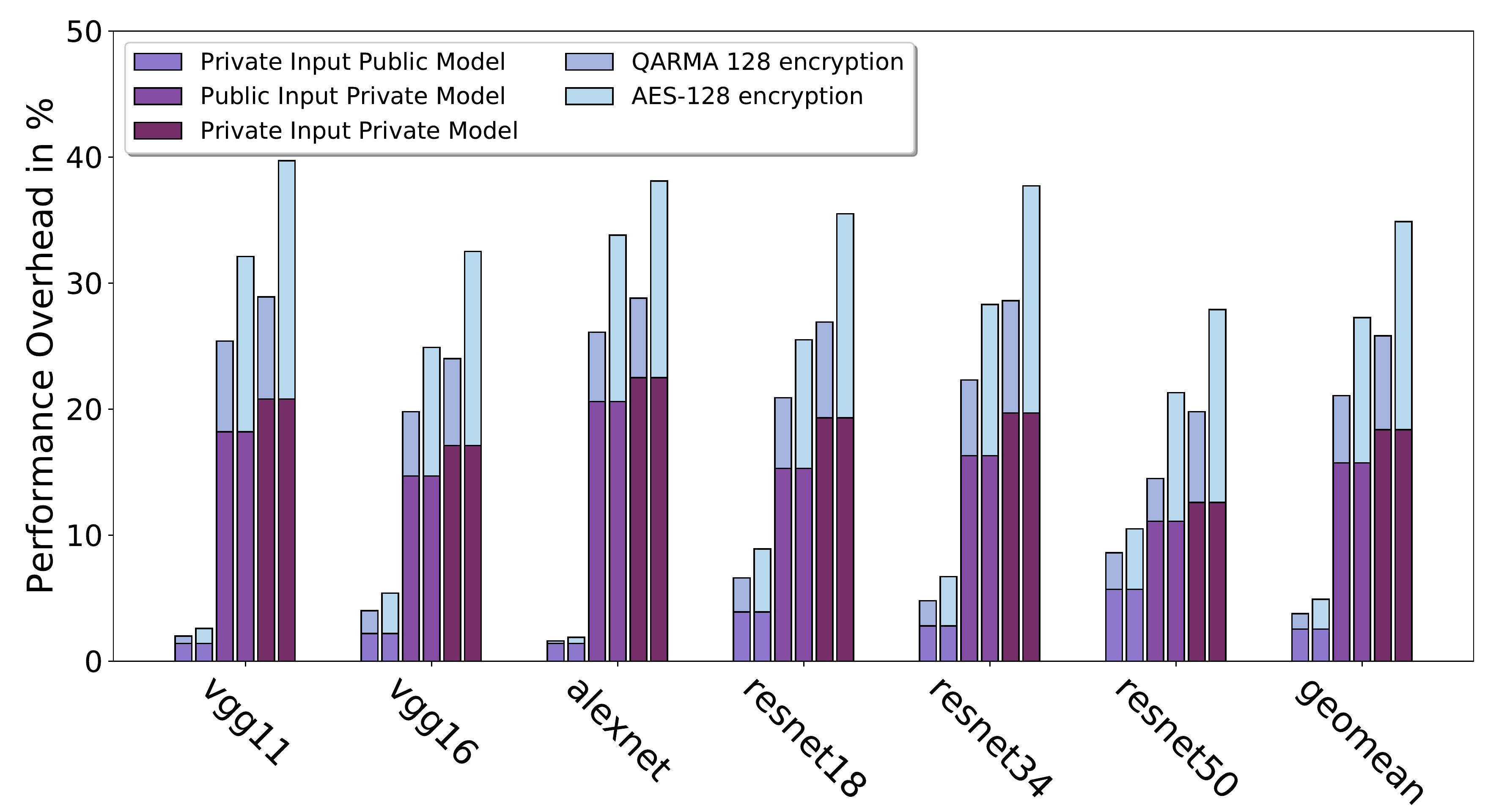} 
        \caption{\label{subfig:temporalSharingPerf}Temporal Sharing}
    \end{subfigure}
    \begin{subfigure}{1.05\linewidth}
        \centering
        \hspace{-1em}
        \includegraphics[width=\linewidth]{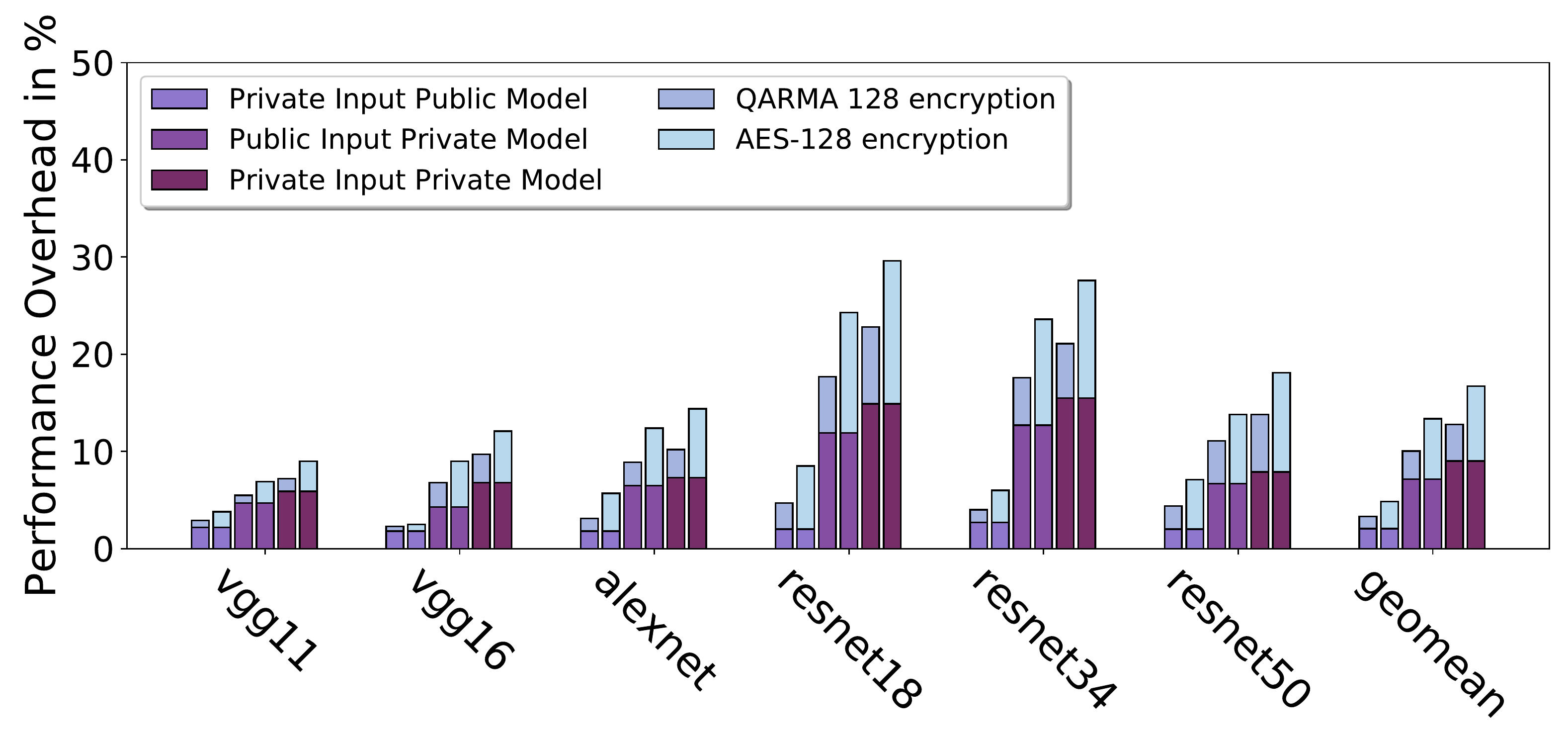} 
        \vspace{-1em}
        \caption{\label{subfig:spatialSharingPerf}Spatial Sharing}
    \end{subfigure}
\caption{\label{fig:perf_overhead}\textbf{Performance Overhead of Two Multi-Tenant Scenarios.
Temporal sharing where multiple tenants share the accelerator in a time-multiplexed manner.
Spatial sharing where resources are evenly shared among four tenants.
Overheads are compared to a non-secure public input public model scenario. Encryption overhead from either QARMA-128 or AES-128 is added on top of each threat model.}}
\vspace{-0.1in}
\end{figure}

\revision{ In this section we study the performance overheads of providing security against the various threat models in both temporal and spatial sharing modes v/s a non-secure baseline for which both model and input are set to be public. Each of the other three threat models has two bars: one for memory encryption defense with QARMA-128 block cipher and the other with AES-128. Private input setting adds \texttt{zeroize} instruction overhead on the input, accumulator and output scratchpad with input padding. Private model further adds overhead due to memory traffic shaper with bandwidth of 400 MB/s, zeroization of weight scratchpad. Figure \ref{subfig:temporalSharingPerf} illustrates that for temporal sharing mode performance overheads range from $3.77\%$ to $25.81\%$ with QARMA128 and $4.92\%$ to $34.87\%$ with AES128 encryption.}

\revision{For spatial sharing (Figure \ref{subfig:spatialSharingPerf}), 
four tenants equally share all accelerator resources including partitioned scratchpad, private queues, execution units and memory traffic bandwidth. In other words, each inference task is given one fourth the resources allocated in temporal sharing mode as shown in Table ~\ref{tbl:SystemSpecifications}. Thus, with spatial sharing, because the non secure baseline is already resource constrained, the relative overheads for providing security are slightly lower. The overhead ranges from $3.31\%$ to $12.78\%$ with QARMA128 and $4.86\%$ to $16.73\%$ with AES128 encryption.The spatial sharing security overhead is less for models with larger layers like alexnet and vgg because limited memory bandwidth in the baseline is primary bottleneck. }

\vspace{0.2in}
\section{Related Work}


ML inference accelerators~\cite{eie,fpga_CNN,acc_survey,VTA,dnnweaver,brainwave} 
extract performance using specialized hardware like DAE\cite{DAE} and systolic array designs\cite{systolic} with domain-specific optimizations\cite{deepcompression}. On the other hand, secure TEEs focus on general-purpose cores~\cite{SGX,bastion_sgx,Sanctum,TZ,keystone} and recently GPUs~\cite{Graviton,telekine} and FPGAs~\cite{hetee} (specifically, on securing the PCIe interface between CPU and GPUs/FPGAs). \system{} is complementary and proposes software-defined enclaves to construct accelerator-TEEs that are tightly coupled on chip with general-purpose cores. 

 
While partitioning\cite{Sanctum,cachePartition,CNNpartition} and shaping\cite{camouflage,mitts} are generic strategies that have been extensively studied from networking to CPU designs, \system{} 
turns them into hardware/RTL modules -- private queues and shapers -- 
to be synthesized based on the threat model 
and tuned based on the program phase information.  
As a result, \system{}'s shaper unit is simpler than memory-controller shapers~\cite{camouflage} because it relies on software to learn traffic distribution and configure it. 
\system{} is also more secure than Camouflage~\cite{camouflage} 
by not assuming that an attacker is limited to observing only coarse-grained signals -- an on-chip co-tenant or privileged software can observe \system{} enclave outputs at arbitrary granularity.  


 

A software configurable \system{} framework is composable with other physical protection units necessary for TEEs like encryption/integrity blocks\cite{mee} and memory access pattern confidentiality using address encryption~\cite{invisimem,obfusmem} or ORAM\cite{PathORM,Phantom}. These memory protection schemes can be exposed to software by extending the \system{} ISA~\cite{mto}. The compiler's auto-tuning framework can be similarly extended to extract performance by performing state-space exploration to (e.g.,) take advantage of streaming patterns for integrity checks.\system{} enables prior work that uses CPU-based TEEs confidential-ML~\cite{Slalom,chiron,Myelin} to use accelerator-TEEs instead.
 
Beyond hardware-based TEEs, cryptographic approaches towards privacy in machine learning have also been widely studied. Homomorphic encryption(HE) and garbled circuits (GC) based research such as Cryptonets\cite{Cryptonets}, Securenets\cite{Securenets},  SecureML\cite{Secureml}, and xonn\cite{Xonn_Raizi} provide confidentiality guarantees for user data without trusting server-side hardware, but are orders of magnitude slower than non-secure execution and do not hide the model. Finally, we observe that protecting against adversarial input attacks (DNNGuard\cite{dnnguard}) is orthogonal to our design and threat model.

\section{Conclusion}
\system{} brings confidential computing to accelerators and introduces {\em software-defined enclaves} -- where the slowdown scales gracefully with the threat model and program phase regularity. The key innovation is to define new hardware modules that replace ubiquitous queues with private queues and adds traffic-shapers at enclave egress -- together, these modules can form a secure \textit{data-plane} for accelerators beyond DAE. While extending \system{} workloads to secure graph programs are near term tasks, bringing these ideas back into general-purpose CPUs and incorporating these modules into hardware verification tools would be the longer term wins. We hope to spur further research into secure and performant enclaves by sharing our code and benchmark suite with the research community.

\bibliographystyle{IEEEtranS}
\bibliography{refs.bib}

\end{document}